%% file: kmbranes1.tex
\input harvmac
\noblackbox

\input epsf
\input tables


\def\journal#1&#2(#3){\unskip, \sl #1\ \bf #2 \rm(19#3) }
\def\andjournal#1&#2(#3){\sl #1~\bf #2 \rm (19#3) }

\def\frac#1#2{{#1\over#2}}

\def\d{\partial}

\def\inbar{\,\vrule height1.5ex width.4pt depth0pt}
\def\IC{\relax\hbox{$\inbar\kern-.3em{\rm C}$}}
\def\IR{\relax{\rm I\kern-.18em R}}
\def\IP{\relax{\rm I\kern-.18em P}}
\def\IZ{\relax{\rm I\kern-.18em Z}}
\def\IN{\relax{\rm I\kenn-.18em N}}

%
%

%
\catcode`\@=11
\def\slash#1{\mathord{\mathpalette\c@ncel{#1}}}
\overfullrule=0pt

\def\HH{{\cal H}}
\def\II{{\cal I}}
\def\JJ{{\cal J}}

\def\LL{{\cal L}}
\def\MM{{\cal M}}
\def\NN{{\cal N}}
\def\OO{{\cal O}}

\def\SS{{\cal S}}

\def\VV{{\cal V}}

\def\ZZ{{\cal Z}}

\def\underrel#1\over#2{\mathrel{\mathop{\kern\z@#1}\limits_{#2}}}

\catcode`\@=12


%

\def\tr{{\rm Tr}}

\def\sinh{{\rm sinh}}
\def\cosh{{\rm cosh}}

\def\exp{{\rm exp}}

\def\ch{{\rm cosh}}



\def\unlockat{\catcode`\@=11}
\def\lockat{\catcode`\@=12}

\unlockat


\def\newsec#1{\global\advance\secno by1\message{(\the\secno. #1)}
\global\subsecno=0\global\subsubsecno=0\eqnres@t\noindent
{\bf\the\secno. #1}
\writetoca{{\secsym} {#1}}\par\nobreak\medskip\nobreak}
\global\newcount\subsecno \global\subsecno=0
\def\subsec#1{\global\advance\subsecno
by1\message{(\secsym\the\subsecno. #1)}
\ifnum\lastpenalty>9000\else\bigbreak\fi\global\subsubsecno=0
\noindent{\it\secsym\the\subsecno. #1}
\writetoca{\string\quad {\secsym\the\subsecno.} {#1}}
\par\nobreak\medskip\nobreak}
\global\newcount\subsubsecno \global\subsubsecno=0
\def\subsubsec#1{\global\advance\subsubsecno by1
\message{(\secsym\the\subsecno.\the\subsubsecno. #1)}
\ifnum\lastpenalty>9000\else\bigbreak\fi
\noindent\quad{\secsym\the\subsecno.\the\subsubsecno.}{#1}
\writetoca{\string\qquad{\secsym\the\subsecno.\the\subsubsecno.}{#1}}
\par\nobreak\medskip\nobreak}

\def\subsubseclab#1{\DefWarn#1\xdef
#1{\noexpand\hyperref{}{subsubsection}%
{\secsym\the\subsecno.\the\subsubsecno}%
{\secsym\the\subsecno.\the\subsubsecno}}%
\writedef{#1\leftbracket#1}\wrlabeL{#1=#1}}
\lockat


\newcount\figno
\figno=0
\def\fig#1#2#3{
\par\begingroup\parindent=0pt\leftskip=1cm\rightskip=1cm\parindent=0pt
\baselineskip=11pt
\global\advance\figno by 1
\midinsert
\epsfxsize=#3
\centerline{\epsfbox{#2}}
\vskip 12pt
{\bf Fig.\ \the\figno: } #1\par
\endinsert\endgroup\par
}
\def\figlabel#1{\xdef#1{\the\figno}}
\def\encadremath#1{\vbox{\hrule\hbox{\vrule\kern8pt\vbox{\kern8pt
\hbox{$\displaystyle #1$}\kern8pt}
\kern8pt\vrule}\hrule}}
%
%


\font\cmss=cmss10
\font\cmsss=cmss10 at 7pt
\def\rlx{\relax\leavevmode}
\def\inbar{\vrule height1.5ex width.4pt depth0pt}
\def\IC{\relax\,\hbox{$\inbar\kern-.3em{\rm C}$}}
\def\IN{\relax{\rm I\kern-.18em N}}
\def\IP{\relax{\rm I\kern-.18em P}}
\def\ZZ{\rlx\leavevmode\ifmmode\mathchoice{\hbox{\cmss Z\kern-.4em Z}}
 {\hbox{\cmss Z\kern-.4em Z}}{\lower.9pt\hbox{\cmsss Z\kern-.36em Z}}
 {\lower1.2pt\hbox{\cmsss Z\kern-.36em Z}}\else{\cmss Z\kern-.4em
 Z}\fi}
\def\IZ{\relax\ifmmode\mathchoice
{\hbox{\cmss Z\kern-.4em Z}}{\hbox{\cmss Z\kern-.4em Z}}
{\lower.9pt\hbox{\cmsss Z\kern-.4em Z}}
{\lower1.2pt\hbox{\cmsss Z\kern-.4em Z}}\else{\cmss Z\kern-.4em
Z}\fi}

\def\narrowplus{\kern -.04truein + \kern -.03truein}
\def\narrowminus{- \kern -.04truein}
\def\narrowminussub{\kern -.02truein - \kern -.01truein}

\def\IZ{\relax\ifmmode\mathchoice
{\hbox{\cmss Z\kern-.4em Z}}{\hbox{\cmss Z\kern-.4em Z}}
{\lower.9pt\hbox{\cmsss Z\kern-.4em Z}}
{\lower1.2pt\hbox{\cmsss Z\kern-.4em Z}}\else{\cmss Z\kern-.4em
Z}\fi}
\def\IB{\relax{\rm I\kern-.18em B}}
\def\IC{{\relax\hbox{$\inbar\kern-.3em{\rm C}$}}}
\def\ID{\relax{\rm I\kern-.18em D}}
\def\IE{\relax{\rm I\kern-.18em E}}
\def\IF{\relax{\rm I\kern-.18em F}}
\def\IG{\relax\hbox{$\inbar\kern-.3em{\rm G}$}}
\def\IGa{\relax\hbox{${\rm I}\kern-.18em\Gamma$}}
\def\IH{\relax{\rm I\kern-.18em H}}
\def\II{\relax{\rm I\kern-.18em I}}
\def\IK{\relax{\rm I\kern-.18em K}}
\def\IP{\relax{\rm I\kern-.18em P}}

\font\cmss=cmss10 \font\cmsss=cmss10 at 7pt
\def\IR{\relax{\rm I\kern-.18em R}}

\def\dab#1{ { \partial \over \partial #1} }

\def\rrangle{{\rangle \rangle}}

\def\brrangle{{\big \rangle \big \rangle}}
\def\bllangle{{\big \langle \big \langle}}
\def\ch{{\rm ch}}
%

%
%
\def\eqnn#1{\xdef #1{(\secsym\the\meqno)}\writedef{#1\leftbracket#1}%
\global\advance\meqno by1\wrlabeL#1}
\def\eqna#1{\xdef #1##1{\hbox{$(\secsym\the\meqno##1)$}}
\writedef{#1\numbersign1\leftbracket#1{\numbersign1}}%
\global\advance\meqno by1\wrlabeL{#1$\{\}$}}
\def\eqn#1#2{\xdef #1{(\secsym\the\meqno)}\writedef{#1\leftbracket#1}%
\global\advance\meqno by1$$#2\eqno#1\eqlabeL#1$$}


\def\boxit#1{\vbox{\hrule\hbox{\vrule\kern8pt
\vbox{\hbox{\kern8pt}\hbox{\vbox{#1}}\hbox{\kern8pt}}
\kern8pt\vrule}\hrule}}
\def\mathboxit#1{\vbox{\hrule\hbox{\vrule\kern5pt\vbox{\kern5pt
\hbox{$\displaystyle #1$}\kern5pt}\kern5pt\vrule}\hrule}}


\lref\Hosomichi{
K.~Hosomichi,
``$\NN=2$ Liouville Theory with Boundary,''
[arXiv:hep-th/0408172].
}

\lref\Fukuda{
T. Fukuda, K. Hosomichi, JHEP 0109 (2001) 003, hep-th/0105217.
}

\lref\DiFrancescoNK{
P.~Di Francesco, P.~Mathieu and D.~Senechal,
``Conformal field theory.''
}

\lref\ZamolodchikovAH{
A.~B.~Zamolodchikov and A.~B.~Zamolodchikov,
``Liouville field theory on a pseudosphere,''
arXiv:hep-th/0101152.
}

\lref\KlebanovYA{
I.~R.~Klebanov and J.~M.~Maldacena,
``Superconformal gauge theories and non-critical superstrings,''
arXiv:hep-th/0409133.
}

\lref\ElitzurPQ{
S.~Elitzur, A.~Giveon, D.~Kutasov, E.~Rabinovici and G.~Sarkissian,
``D-branes in the background of NS fivebranes,''
JHEP {\bf 0008}, 046 (2000)
[arXiv:hep-th/0005052].
}

\lref\BoucherBH{
W.~Boucher, D.~Friedan and A.~Kent,
``Determinant Formulae And Unitarity For The N=2 Superconformal Algebras In
Two-Dimensions Or Exact Results On String Compactification,''
Phys.\ Lett.\ B {\bf 172}, 316 (1986).
}

\lref\NamHU{
S.~k.~Nam,
``Superconformal And Super Kac-Moody Invariant Quantum Field Theories In
Two-Dimensions,''
Phys.\ Lett.\ B {\bf 187}, 340 (1987).
}

\lref\KiritsisRV{
E.~Kiritsis,
``Character Formulae And The Structure Of The Representations Of The N=1, N=2
Superconformal Algebras,''
Int.\ J.\ Mod.\ Phys.\ A {\bf 3}, 1871 (1988).
}

\lref\KazamaQP{
Y.~Kazama and H.~Suzuki,
``New N=2 Superconformal Field Theories And Superstring Compactification,''
Nucl.\ Phys.\ B {\bf 321}, 232 (1989).
}

\lref\AharonyXN{
O.~Aharony, A.~Giveon and D.~Kutasov,
``LSZ in LST,''
Nucl.\ Phys.\ B {\bf 691}, 3 (2004)
[arXiv:hep-th/0404016].
}

\lref\WittenYR{
E.~Witten,
``On string theory and black holes,''
Phys.\ Rev.\ D {\bf 44}, 314 (1991).
}

\lref\DijkgraafBA{
R.~Dijkgraaf, H.~Verlinde and E.~Verlinde,
``String propagation in a black hole geometry,''
Nucl.\ Phys.\ B {\bf 371}, 269 (1992).
}

\lref\HoriAX{
K.~Hori and A.~Kapustin,
``Duality of the fermionic 2d black hole and N = 2 Liouville theory as  mirror
symmetry,''
JHEP {\bf 0108}, 045 (2001)
[arXiv:hep-th/0104202].
}

\lref\DobrevHQ{
V.~K.~Dobrev,
``Characters Of The Unitarizable Highest Weight Modules Over The N=2
Superconformal Algebras,''
Phys.\ Lett.\ B {\bf 186}, 43 (1987).
}

\lref\EguchiIK{
T.~Eguchi and Y.~Sugawara,
``Modular bootstrap for boundary N = 2 Liouville theory,''
JHEP {\bf 0401}, 025 (2004)
[arXiv:hep-th/0311141].
}

\lref\EguchiTC{
T.~Eguchi and Y.~Sugawara,
``Modular invariance in superstring on Calabi-Yau n-fold with A-D-E
singularity,''
Nucl.\ Phys.\ B {\bf 577}, 3 (2000)
[arXiv:hep-th/0002100].
}

\lref\MizoguchiKK{
S.~Mizoguchi,
``Modular invariant critical superstrings on four-dimensional Minkowski  space
x two-dimensional black hole,''
JHEP {\bf 0004}, 014 (2000)
[arXiv:hep-th/0003053].
}

\lref\MurthyES{
S.~Murthy,
``Notes on non-critical superstrings in various dimensions,''
JHEP {\bf 0311}, 056 (2003)
[arXiv:hep-th/0305197].
}

\lref\KutasovPV{
D.~Kutasov,
``Some properties of (non)critical strings,''
arXiv:hep-th/9110041.
}

\lref\KutasovUA{
D.~Kutasov and N.~Seiberg,
``Noncritical Superstrings,''
Phys.\ Lett.\ B {\bf 251}, 67 (1990).
}

\lref\HananyEV{
A.~Hanany, N.~Prezas and J.~Troost,
``The partition function of the two-dimensional black hole conformal  field
theory,''
JHEP {\bf 0204}, 014 (2002)
[arXiv:hep-th/0202129].
}

\lref\IsraelIR{
D.~Israel, C.~Kounnas, A.~Pakman and J.~Troost,
``The partition function of the supersymmetric two-dimensional black hole and
little string theory,''
JHEP {\bf 0406}, 033 (2004)
[arXiv:hep-th/0403237].
}

\lref\EguchiYI{
T.~Eguchi and Y.~Sugawara,
``SL(2,R)/U(1) supercoset and elliptic genera of non-compact Calabi-Yau
manifolds,''
JHEP {\bf 0405}, 014 (2004)
[arXiv:hep-th/0403193].
}

\lref\FotopoulosUT{
A.~Fotopoulos, V.~Niarchos and N.~Prezas,
``D-branes and extended characters in SL(2,R)/U(1),''
Nucl.\ Phys.\ B {\bf 710}, 309 (2005)
[arXiv:hep-th/0406017].
}

\lref\EguchiIKK{
T.~Eguchi and Y.~Sugawara,
``Conifold type singularities, N = 2 Liouville and SL(2:R)/U(1) theories,''
arXiv:hep-th/0411041.
}

\lref\DiVecchiaRH{
P.~Di Vecchia and A.~Liccardo,
``D branes in string theory. I,''
NATO Adv.\ Study Inst.\ Ser.\ C.\ Math.\ Phys.\ Sci.\  {\bf 556}, 1 (2000)
[arXiv:hep-th/9912161].
}

\lref\DiVecchiaFX{
P.~Di Vecchia and A.~Liccardo,
``D-branes in string theory. II,''
arXiv:hep-th/9912275.
}

\lref\GaberdielJR{
M.~R.~Gaberdiel,
``Lectures on non-BPS Dirichlet branes,''
Class.\ Quant.\ Grav.\  {\bf 17}, 3483 (2000)
[arXiv:hep-th/0005029].
}

\lref\OoguriCK{
  H.~Ooguri, Y.~Oz and Z.~Yin,
  ``D-branes on Calabi-Yau spaces and their mirrors,''
  Nucl.\ Phys.\ B {\bf 477}, 407 (1996)
  [arXiv:hep-th/9606112].
}

\lref\GiveonPX{
A.~Giveon and D.~Kutasov,
``Little string theory in a double scaling limit,''
JHEP {\bf 9910}, 034 (1999)
[arXiv:hep-th/9909110].
}

\lref\FZZ{
V.~A.~Fateev, A.~B.~Zamolodchikov and Al.~B.~Zamolodchikov, unpublished.
}

\lref\AharonyUB{
O.~Aharony, M.~Berkooz, D.~Kutasov and N.~Seiberg,
``Linear dilatons, NS5-branes and holography,''
JHEP {\bf 9810}, 004 (1998)
[arXiv:hep-th/9808149].
}

\lref\GiveonZM{
A.~Giveon, D.~Kutasov and O.~Pelc,
``Holography for non-critical superstrings,''
JHEP {\bf 9910}, 035 (1999)
[arXiv:hep-th/9907178].
}

\lref\KutasovUF{
D.~Kutasov,
``Introduction to little string theory,''
{\it Prepared for ICTP Spring School on Superstrings and Related Matters, Trieste, Italy, 2-10 Apr 2001}
}

\lref\AharonyKS{
O.~Aharony,
``A brief review of 'little string theories',''
Class.\ Quant.\ Grav.\  {\bf 17}, 929 (2000)
[arXiv:hep-th/9911147].
}

\lref\GiveonSR{
A.~Giveon and D.~Kutasov,
``Brane dynamics and gauge theory,''
Rev.\ Mod.\ Phys.\  {\bf 71}, 983 (1999)
[arXiv:hep-th/9802067].
}

\lref\PolyakovJU{
A.~M.~Polyakov,
``The wall of the cave,''
Int.\ J.\ Mod.\ Phys.\ A {\bf 14}, 645 (1999)
[arXiv:hep-th/9809057].
}

\lref\ElitzurPQ{
S.~Elitzur, A.~Giveon, D.~Kutasov, E.~Rabinovici and G.~Sarkissian,
``D-branes in the background of NS fivebranes,''
JHEP {\bf 0008}, 046 (2000)
[arXiv:hep-th/0005052].
}

\lref\FateevIK{
V.~Fateev, A.~B.~Zamolodchikov and A.~B.~Zamolodchikov,
``Boundary Liouville field theory. I: Boundary state and boundary  two-point
function,''
arXiv:hep-th/0001012.
}

\lref\TeschnerMD{
J.~Teschner,
``Remarks on Liouville theory with boundary,''
arXiv:hep-th/0009138.
}

\lref\ZamolodchikovAH{
A.~B.~Zamolodchikov and A.~B.~Zamolodchikov,
``Liouville field theory on a pseudosphere,''
arXiv:hep-th/0101152.
}

\lref\RibaultSS{
S.~Ribault and V.~Schomerus,
``Branes in the 2-D black hole,''
JHEP {\bf 0402}, 019 (2004)
[arXiv:hep-th/0310024].
}

\lref\EguchiIK{
T.~Eguchi and Y.~Sugawara,
``Modular bootstrap for boundary N = 2 Liouville theory,''
JHEP {\bf 0401}, 025 (2004)
[arXiv:hep-th/0311141].
}

\lref\AhnTT{
C.~Ahn, M.~Stanishkov and M.~Yamamoto,
``One-point functions of N = 2 super-Liouville theory with boundary,''
Nucl.\ Phys.\ B {\bf 683}, 177 (2004)
[arXiv:hep-th/0311169].
}

\lref\IsraelJT{
D.~Israel, A.~Pakman and J.~Troost,
``D-branes in N = 2 Liouville theory and its mirror,''
arXiv:hep-th/0405259.
}

\lref\AhnQB{
C.~Ahn, M.~Stanishkov and M.~Yamamoto,
``ZZ-branes of N = 2 super-Liouville theory,''
JHEP {\bf 0407}, 057 (2004)
[arXiv:hep-th/0405274].
}

\lref\HosomichiPH{
K.~Hosomichi,
``N = 2 Liouville theory with boundary,''
arXiv:hep-th/0408172.
}

\lref\IsraelFN{
D.~Israel, A.~Pakman and J.~Troost,
``D-branes in little string theory,''
arXiv:hep-th/0502073.
}

\lref\progress{
A.~Fotopoulos, V.~Niarchos and N.~Prezas,
work in progress.
}

\lref\KlemmVN{
H.~Klemm,
``Embedding diagrams of the N = 2 superconformal algebra under spectral
flow,''
Int.\ J.\ Mod.\ Phys.\ A {\bf 19}, 5263 (2004)
[arXiv:hep-th/0306073].
}

\lref\GatoRiveraMI{
B.~Gato-Rivera,
``Recent results on N = 2 superconformal algebras,''
arXiv:hep-th/0002081.
}

\lref\dorrzapf{
M.~D\"orrzapf,
``Superconformal Field Theories and Their Representations,'' Ph.D.~ Thesis,
University of Cambridge, 1995.
}

\lref\MaldacenaHW{
J.~M.~Maldacena and H.~Ooguri,
``Strings in AdS(3) and SL(2,R) WZW model. I,''
J.\ Math.\ Phys.\  {\bf 42}, 2929 (2001)
[arXiv:hep-th/0001053].
}

\lref\BilalUH{
A.~Bilal and J.~L.~Gervais,
``New Critical Dimensions For String Theories,''
Nucl.\ Phys.\ B {\bf 284}, 397 (1987).
}

\lref\BilalIA{
A.~Bilal and J.~L.~Gervais,
``Modular Invariance For Closed Strings At The New Critical Dimensions,''
Phys.\ Lett.\ B {\bf 187}, 39 (1987).
}

\lref\PonsotGT{
B.~Ponsot, V.~Schomerus and J.~Teschner,
``Branes in the Euclidean AdS(3),''
JHEP {\bf 0202}, 016 (2002)
[arXiv:hep-th/0112198].
}

\lref\OoguriIH{
H.~Ooguri and C.~Vafa,
``Geometry of N = 1 dualities in four dimensions,''
Nucl.\ Phys.\ B {\bf 500}, 62 (1997)
[arXiv:hep-th/9702180].
}

\lref\SeibergPQ{
N.~Seiberg,
``Electric - magnetic duality in supersymmetric nonAbelian gauge theories,''
Nucl.\ Phys.\ B {\bf 435}, 129 (1995)
[arXiv:hep-th/9411149].
}

\lref\HananySA{
A.~Hanany and A.~Zaffaroni,
``Chiral symmetry from type IIA branes,''
Nucl.\ Phys.\ B {\bf 509}, 145 (1998)
[arXiv:hep-th/9706047].
}

\lref\BrodieSZ{
J.~H.~Brodie and A.~Hanany,
``Type IIA superstrings, chiral symmetry, and N = 1 4D gauge theory
dualities,''
Nucl.\ Phys.\ B {\bf 506}, 157 (1997)
[arXiv:hep-th/9704043].
}

\lref\BertoliniJY{
M.~Bertolini, P.~Di Vecchia, M.~Frau, A.~Lerda, R.~Marotta and R.~Russo,
``Is a classical description of stable non-BPS D-branes possible?,''
Nucl.\ Phys.\ B {\bf 590}, 471 (2000)
[arXiv:hep-th/0007097].
}

\lref\ElitzurHC{
S.~Elitzur, A.~Giveon, D.~Kutasov, E.~Rabinovici and A.~Schwimmer,
``Brane dynamics and N = 1 supersymmetric gauge theory,''
Nucl.\ Phys.\ B {\bf 505}, 202 (1997)
[arXiv:hep-th/9704104].
}

\lref\WittenSC{
E.~Witten,
``Solutions of four-dimensional field theories via M-theory,''
Nucl.\ Phys.\ B {\bf 500}, 3 (1997)
[arXiv:hep-th/9703166].
}

\lref\GiveonMI{
A.~Giveon, D.~Kutasov, E.~Rabinovici and A.~Sever,
``Phases of quantum gravity in AdS(3) and linear dilaton backgrounds,''
arXiv:hep-th/0503121.
}

\lref\BarsSR{
I.~Bars and K.~Sfetsos,
``Conformally exact metric and dilaton in string theory on curved
space-time,''
Phys.\ Rev.\ D {\bf 46}, 4510 (1992)
[arXiv:hep-th/9206006].
}

\lref\TseytlinMY{
A.~A.~Tseytlin,
``Conformal sigma models corresponding to gauged Wess-Zumino-Witten
theories,''
Nucl.\ Phys.\ B {\bf 411}, 509 (1994)
[arXiv:hep-th/9302083].
}

\lref\FotopoulosVC{
A.~Fotopoulos,
``Semiclassical description of D-branes in SL(2)/U(1) gauged WZW model,''
Class.\ Quant.\ Grav.\  {\bf 20}, S465 (2003)
[arXiv:hep-th/0304015].
}

\lref\AlishahihaYV{
M.~Alishahiha, A.~Ghodsi and A.~E.~Mosaffa,
``On isolated conformal fixed points and noncritical string theory,''
JHEP {\bf 0501}, 017 (2005)
[arXiv:hep-th/0411087].
}

\lref\KupersteinYF{
S.~Kuperstein and J.~Sonnenschein,
``Non-critical, near extremal AdS(6) background as a holographic laboratory
of four dimensional YM theory,''
JHEP {\bf 0411}, 026 (2004)
[arXiv:hep-th/0411009].
}

\lref\ItohMT{
K.~Itoh, H.~Kunitomo, N.~Ohta and M.~Sakaguchi,
``BRST Analysis of physical states in two-dimensional black hole,''
Phys.\ Rev.\ D {\bf 48}, 3793 (1993)
[arXiv:hep-th/9305179].
}

\lref\ElitzurFH{
S.~Elitzur, A.~Giveon and D.~Kutasov,
``Branes and N = 1 duality in string theory,''
Phys.\ Lett.\ B {\bf 400}, 269 (1997)
[arXiv:hep-th/9702014].
}



\Title{
{\vbox{
\rightline{CPHT-RR021.0305}
\vskip .5pt
\rightline{CRE-TH-05/06}
\vskip .5pt
\rightline{NEIP-05-04}
}}}
{\vbox{\centerline{D-branes and SQCD}
\vskip 10pt \centerline{In Non-Critical Superstring Theory}
}}
\centerline{Angelos Fotopoulos$^a$\foot{Also at the Centre de Physique
Theorique, Ecole Polytechnique, Palaiseau, 91128, France.}
, Vasilis Niarchos$^b$ and Nikolaos Prezas$^c$}
\medskip
\centerline{{\it $^a$ Department of Physics, University of Crete}}
\centerline{{\it 710 03 Heraklion, Greece}}
\smallskip
\centerline{{\it $^b$ The Niels Bohr Institute}}
\centerline{\it Blegdamsvej 17, 2100 Copenhagen \O, Denmark}
\smallskip
\centerline{{\it $^c$ Institut de Physique, Universit\'e de Neuch\^atel}}
\centerline{{\it CH--2000 Neuch\^atel, Switzerland}}
\medskip\bigskip
\noindent

Using exact boundary conformal field theory methods
we analyze the D-brane physics of a specific four-dimensional
non-critical superstring theory which involves the $\NN=2$
$SL(2)/U(1)$ Kazama-Suzuki model at level $1$. Via the holographic duality
of \GiveonZM\ our results are relevant for D-brane dynamics
in the background of NS5-branes and D-brane dynamics near a conifold singularity.
We pay special attention to a configuration of
D$3$- and D$5$-branes that realizes $\NN=1$ supersymmetric QCD and discuss the
massless spectrum and classical moduli of this setup in detail.
We also comment briefly on the implications of this construction for the recently
proposed generalization of the AdS/CFT correspondence by Klebanov
and Maldacena within the setting of non-critical superstrings.

\vfill
\Date{April, 2005}


\listtoc
\writetoc

\newsec{Introduction}

Non-critical superstring theories \refs{\KutasovUA,\KutasovPV} can be
formulated in $d=2n$ ($n=0,\ldots,4$)\foot{$n=4$ is the critical
ten-dimensional fermionic string.}
spacetime dimensions and describe fully consistent
solutions of string theory in subcritical dimensions. They have $\NN=(2,2)$
worldsheet supersymmetry and appropriate spacetime supersymmetry
consisting of (at least) $2^{n+1}$ spacetime supercharges.
On the worldsheet, these theories typically develop a dynamical Liouville mode
and they have a target space of the form
\eqn\iaa{
\IR^{d-1,1} \times \IR_{\phi} \times S^1 \times \MM
~,}
where $\IR_{\phi}$ is a linear dilaton direction,
$S^1$ is a compact boson and $\MM$ is described by a
worldsheet theory with $\NN=2$ supersymmetry, $e.g.$
a Landau-Ginzburg theory or a Gepner product thereof.
Due to the linear dilaton, these theories have a
strong coupling singularity, which can be resolved
in two equivalent ways:
\item{(1)} We can add to the worldsheet Lagrangian
a superpotential term of the following form (in superspace
language):
\eqn\iab{
\delta \LL = \mu \int d^2 z d^2 \theta e^{-\frac{1}{Q}(\phi+iY)}
+ c.c.
~}
$Q$ denotes the linear dilaton slope, $\phi$ parametrizes
the linear dilaton direction and $Y$ parametrizes the $S^1$.
This interaction couples the $\IR_{\phi}$ and $S^1$
theories into the well-known $\NN=2$ Liouville theory.
\item{(2)} An alternative way to resolve the strong coupling
singularity can be achieved by replacing the $\IR_{\phi}\times S^1$
part of the background \iaa\ with the $\NN=2$ Kazama-Suzuki
supercoset $SL(2)_k/U(1)$ at level $k=2/Q^2$.
This space has a cigar-shaped geometry and provides a geometric
cut-off for the strong coupling singularity.

\noindent
The $\NN=2$ Liouville theory and the $\NN=2$
Kazama-Suzuki model are known to be equivalent by mirror-symmetry.
This non-trivial statement is the supersymmetric version
of a similar conjecture in the bosonic case \FZZ\ involving
the Sine-Liouville theory and the bosonic $SL(2)/U(1)$ theory.
The supersymmetric extension was
first conjectured in \GiveonPX\ and later proven in \HoriAX.

Non-critical superstring theories are interesting for
a number of reasons. First of all, it has been argued on general
grounds \AharonyUB\ that theories with asymptotic linear dilaton directions
are holographic. In particular, \GiveonZM\
found that the holographic dual of the
$d$-dimensional theory \iaa\ is a corresponding $d$-dimensional
Little String Theory (LST) (for a review see \refs{\AharonyKS, \KutasovUF}).
LST's are non-local, non-gravitational interacting  theories
that can be defined by taking suitable scaling limits on the worldvolume
of NS5-branes or in critical string theory near Calabi-Yau singularities.

LST's appear in various applications. The one that
will be the focal point of this paper
involves four-dimensional gauge theories that can be realized
on D-branes stretched between NS5-branes (for a review
of the subject see \GiveonSR). A typical brane configuration
that realizes four-dimensional $\NN=1$ super-Yang-Mills (SYM),
say in type IIA string theory, consists of two NS5-branes
and $N_c$ D4-branes oriented as follows (see fig.\ 1):
\eqn\iac{\eqalign{
NS5: ~ &(x^0,x^1,x^2,x^3,x^4,x^5)
\cr
NS5': ~ &(x^0,x^1,x^2,x^3,x^8,x^9)
\cr
D4: ~ &(x^0,x^1,x^2,x^3,x^6)
~}}
The NS5-branes are tilted with respect to each other
breaking supersymmetry by one quarter.
The $N_c$ D4-branes stretched between the NS5-branes along the $6$-direction
break the overall supersymmetry by an additional one-half and realize
a gauge theory with four supercharges and gauge group
$U(N_c)$.

\bigskip
{\vbox{{\epsfxsize=90mm
\nobreak
\centerline{\epsfbox{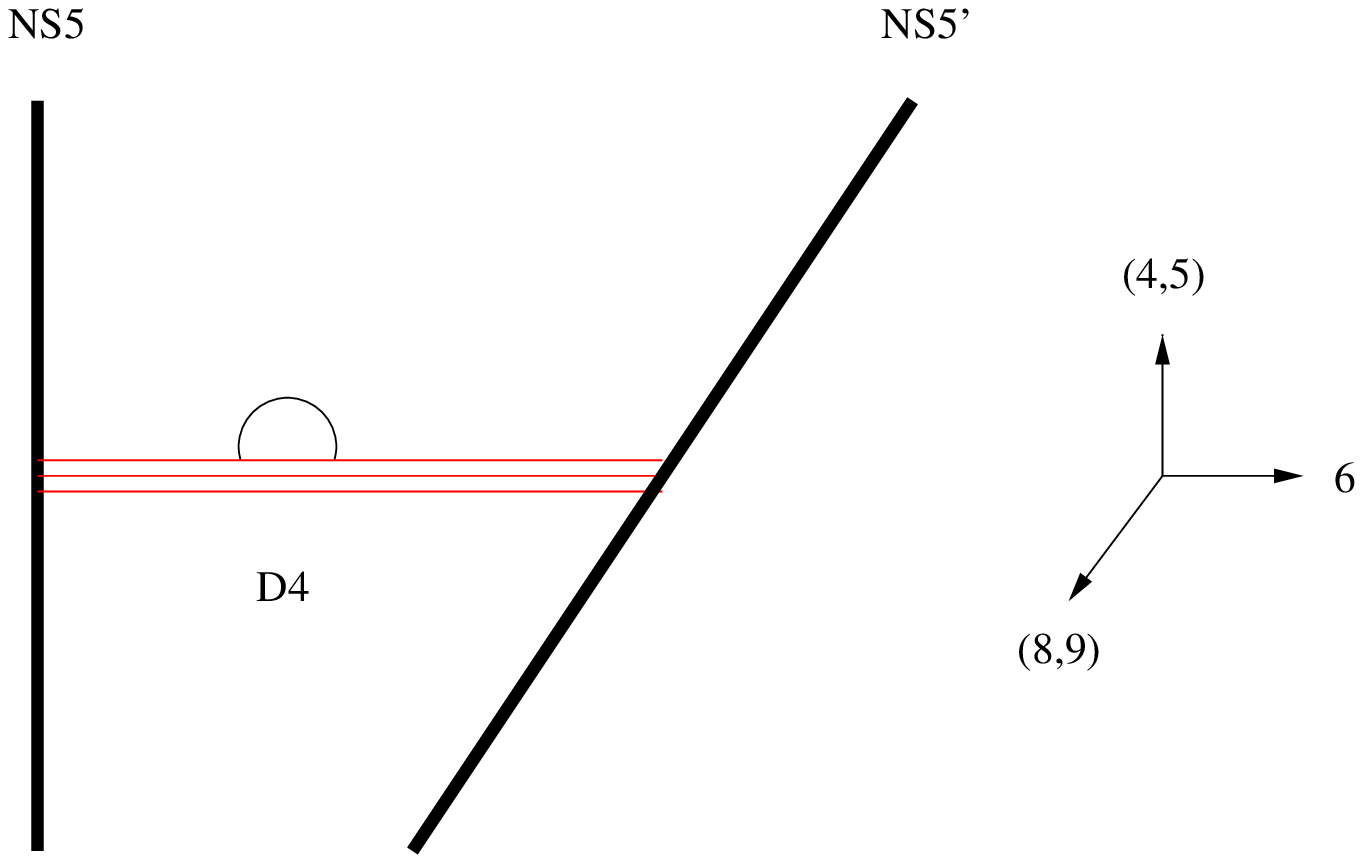}}
\nobreak\bigskip
{\raggedright\it \vbox{
{\bf Figure 1.}
{\it A configuration of two NS5-branes and $N_c$ suspended D4-branes
that realizes $\NN=1$ SYM. Flavors can be introduced by adding appropriately
oriented D6-branes  or semi-infinite D4-branes.
} }}}}
\bigskip}

In order to obtain a truly four-dimensional gauge theory and to decouple
the gauge dynamics from other complications of string theory
we need to take the double-scaling limit
\eqn\iad{
g_s \rightarrow 0~, ~ ~ L\rightarrow 0~, ~ ~
g_{\rm YM}^2=\frac{g_s l_s}{L}={\rm fixed}
~,}
where $L$ is the length of the finite D4-branes in
the 6-direction and the limit is taken in such a way
that the effective $g_{YM}$ coupling of the gauge theory
is kept fixed. This limit is the same as the double
scaling limit of LST \GiveonPX\ and, via the holographic duality of
\GiveonZM, the same brane configuration can be realized by taking
$N_c$ D3-branes in the non-critical superstring theory
\eqn\iae{
\IR^{3,1}\times SL(2)_1/U(1)
~.}
The D3-branes are extended in $\IR^{3,1}$
and are localized near the tip of the cigar-shaped target space of
$SL(2)_1/U(1)$. Flavors can also be realized in this
setup by adding D4- or D5-branes in \iae\ (see below for explicit
constructions). Equivalently, in the original brane configuration of fig.\ 1
flavors can be introduced by
adding appropriately oriented D4- or D6-branes
(see $e.g.$ \GiveonSR\ or fig.\ 4 in section 5 below).

The main purpose of this paper is to analyze the physics of
such D-brane configurations in the non-critical superstring
\iae\ using exact boundary conformal field theory methods.
Similar configurations of D-branes in type IIB non-critical string theory
have been considered recently by Klebanov and Maldacena \KlebanovYA.
The authors of that paper analyzed a configuration
of D3-, D5-, and anti-D5-branes\foot{The presence of anti-D5-branes
in \KlebanovYA\ was anticipated on the basis of certain tadpole cancellation
conditions. In what follows, we argue that such conditions are automatically satisfied
for the D5-branes we formulate and there is no need to introduce anti-D5-branes.}
in 6-dimensional supergravity and proposed a very interesting generalization of the AdS/CFT
correspondence within the context of non-critical superstrings.\foot{For relevant discussions
and follow-up work in this direction see \refs{\KupersteinYF,\AlishahihaYV}.}
The supergravity results pointed towards an $AdS_5 \times S^1$ holographic
dual of $\NN=1$ SQCD in the conformal window.
The present work adds a different element to this story by analyzing
the relevant D3/D5 configuration from the open
string theory point of view. This is bound to be useful for analyzing
further aspects of the proposed holographic duality.
In general, the connection between non-critical strings
and four-dimensional gauge theories has long been anticipated \PolyakovJU\
and we hope that the present analysis will be relevant for
similar investigations of gauge theories in related contexts.

We should mention that a closely related analysis of D-branes
in the background of NS5-branes has been performed
previously in \ElitzurPQ\ . This paper analyzed
various aspects of the dynamics of D6-branes and semi-infinite D4-branes
in the near horizon geometry of NS5-branes with the use of worldsheet
techniques and verified several of the expected properties
of the gauge theories realized in this setting.
Due to important recent progress in the study of the boundary
conformal field theory of $SL(2)/U(1)$
\refs{\RibaultSS\EguchiIK\AhnTT\IsraelJT\AhnQB\FotopoulosUT-\HosomichiPH},
motivated by the seminal work of
\refs{\FateevIK, \TeschnerMD, \ZamolodchikovAH},
we are now in position to discuss some additional aspects of this story.
Most notably, we have a better control on the properties of the D0-branes
localized near the tip of the cigar, which lead to
the $finite$ D4-branes of fig.\ 1. Indeed, we will see how the technology
of \refs{\RibaultSS\EguchiIK\AhnTT\IsraelJT\AhnQB\FotopoulosUT-\HosomichiPH}
yields the full spectrum of open strings stretching on such branes
and how we can use it to engineer interesting QCD-like theories.
A related analysis of D-branes in the background of NS5-branes using
similar techniques
has appeared recently in \IsraelFN.

The layout of this paper is as follows.
In section 2, we review the basic characteristics of type $0$ and
type II non-critical superstring theory on \iae, establish our notation
and summarize the key features of the closed string spectrum.
In section 3, we proceed to analyze the D-brane physics of the theory by
using boundary conformal field theory methods,
which allow for explicit computations of the cylinder
amplitudes and open string spectra. Adapting the existing knowledge
on $SL(2)/U(1)$ D-branes in the current setup we obtain BPS and
non-BPS D3-, D4- and D5-branes and discuss their properties.
For simplicity, we focus on D-branes with Neumann boundary
conditions in all four flat directions of \iae. In section 4,
we discuss general properties of the
BPS D3- and D5-branes of the type IIB theory. We are especially
interested in the massless RR couplings of these branes and the presence
(or absence) of potential tadpole cancellation conditions.
This sets the stage for the main purpose of this paper; the realization
of $\NN=1$ SQCD theories on appropriate D-brane setups within the non-critical
superstring theory. In section 5 we show explicitly, how this can be achieved
with a particular D3-D5 setup that realizes the electric description of
$\NN=1$ SQCD. Also, we compare the classical symmetries and
moduli of the D-brane configuration with those expected from the
gauge theory and find agreement as in previous investigations
of this subject \GiveonSR. In this discussion the Higgsing moduli
and the ability (or inability) to formulate the magnetic description
of $\NN=1$ SQCD are particularly interesting points, which
appear to be alluding to some yet unexplored properties of D-branes
on $SL(2)/U(1)$. We conclude in section 6 with a brief discussion
of our results and interesting future prospects related
to Seiberg duality and the holographic duality proposed
in \KlebanovYA. Two appendices contain useful
information about the properties of the
$SL(2)/U(1)$ characters and the GSO projected torus partition sum
of the four-dimensional non-critical superstring theory.

\newsec{Non-critical superstrings}

In this section we review the most prominent features of the closed
string sector of the four-dimensional
non-critical superstring theory that we want to analyze, establish
our notation and present the torus partition function of the type $0$
and type II theories.

\subsec{Notation and representation content of the $SL(2)/U(1)$ supercoset}

The non-trivial part of the worldsheet theory with target space \iaa\ is
the two-dimensional superconformal theory $SL(2)_k/U(1)$ \KazamaQP.
This theory can be obtained from the supersymmetric $SL(2,\IR)$ WZW model
at level $k$ by gauging an appropriate $U(1)$ subgroup (the details
of this gauging can be found in various references - see, for example
\AharonyXN). It has $\NN=(2,2)$ worldsheet supersymmetry and central charge
\eqn\aaa{
\hat c=\frac{c}{3}=1+\frac{2}{k}
~.}
In general, $k$ can be any positive real number but
in this paper we set $k=1$.\foot{The cases with $k>1$
and $k<1$ exhibit interesting differences. See \GiveonMI\
for a recent discussion.} We want
to couple $SL(2)_k/U(1)$ to four-dimensional Minkowski
space to obtain a Weyl-anomaly free fermionic string. This
implies that the total central charge has to be 15, $i.e.$
\eqn\aab{
c_{\rm flat}+c_{\rm coset}=15 \Leftrightarrow k=1
~.}

As a sigma-model, $SL(2)/U(1)$ describes string propagation on a
cigar-shaped two-dimensional manifold
\refs{\WittenYR,\DijkgraafBA} with metric \eqn\aac{
ds^2=k(d\rho^2+\tanh^2\rho d\theta^2)~, ~ ~ \theta \sim
\theta+2\pi ~,} vanishing $B$-field and varying dilaton \eqn\aad{
\Phi(\rho)=-\log \cosh\rho+\Phi_0 ~.}
This background receives
$\alpha'$ corrections in the bosonic case \DijkgraafBA, but is
exact in the supersymmetric case \refs{\BarsSR, \TseytlinMY},
which is the case of interest in this paper. The value of the
dilaton $\Phi_0$ at the tip of the cigar is a free tunable
parameter. T-duality along the angular direction of the cigar acts
non-trivially and the resulting geometry, which naively looks like
a trumpet, is described by a closely related $\NN=(2,2)$
superconformal field theory - the $\NN=2$ Liouville theory
\HoriAX.

The representation theory of $SL(2)/U(1)$ is a useful tool
for the analysis of the closed string spectrum and the formulation
of D-branes on the cigar geometry \aac, \aad. Since we use it
heavily in later sections, it is a good idea to review here the
basic unitary representations of $SL(2)/U(1)$ and the corresponding
characters. This will also set up our notation.
The representations are labeled by the
scaling dimension $h$ and the $U(1)_R$-charge $Q$.
The $unitary$ highest-weight representations of the $\NN=2$
Kazama-Suzuki model fall into the following three classes
\refs{\DobrevHQ,\KiritsisRV,\ItohMT}:\foot{The representation theory
of the $\NN=2$ superconformal algebra is an interesting subject on its own
\refs{\DobrevHQ, \KiritsisRV, \dorrzapf, \GatoRiveraMI, \KlemmVN}.
In certain cases, $\NN=2$ representations exhibit more involved
embedding diagrams associated with the appearance of ``sub-singular''
vectors and the computation of the corresponding characters
becomes highly non-trivial. It is commonly believed however that
the unitary representations presented here do not suffer from these subtleties.
We would like to thank T.\ Eguchi, M.\ Gaberdiel, E.\ Kiritsis, H.\ Klemm and
Y.\ Sugawara for helpful correspondence on these issues.}
\item{($a$)} $Continuous$ $representations$:
These are non-degenerate representations with
\eqn\aae{
h_{j,m}=\frac{-j(j-1)+m^2}{k}~, ~ ~ Q_m=\frac{2m}{k}
~,}
and
\eqn\aaf{
j=\frac{1}{2}+is~, ~ ~ s\in \IR_{\geq 0}~, ~ ~ m=r+\alpha~, ~ r\in \IZ~, ~
\alpha\in [0,1)
~.}
The NS-sector characters read:\foot{The  $\widetilde{NS}$-, $R$-
and $\widetilde{R}$-sector characters will be presented below.}
\eqn\aag{
\ch_c(h_{j,m},Q_m;\tau,z)\bigg[ {0 \atop 0}\bigg]=
q^{h_{j,m}-(\hat c-1)/8} y^{Q_m}
\frac{\theta \big[ {0 \atop 0}\big] (\tau,z)}{\eta(\tau)^3}
~,}
where as usual we set $q=e^{2\pi i\tau}$ and $y=e^{2\pi i z}$.
$\theta\big[ {a \atop b }\big](\tau,z)$, with $a,b=0,1$, are
the standard $\theta$-functions whose properties we summarize
in appendix A.
\item{($b$)} $Discrete$ $representations$:
These are degenerate representations with\foot{This unitarity bound
is restricted further in physical theories to $\frac{1}{2}<j<\frac{k+1}{2}$
\refs{\GiveonPX, \MaldacenaHW, \HananyEV}.}
\eqn\aai{
j \in \IR~, ~ ~ 0<j<\frac{k+2}{2}~,~ ~ r\in \IZ
~}
and
\eqn\aaj{
h_{j,r}=\frac{-j(j-1)+(j+r)^2}{k}~,
~ ~ Q_{j+r}=\frac{2(j+r)}{k}~, ~ ~ r\geq 0
~,}
\eqn\aak{
h_{j,r}=\frac{-j(j-1)+(j+r)^2}{k}-r-\frac{1}{2}~,
~ ~ Q_{j,r}=\frac{2(j+r)}{k}-1
~, ~ ~ r<0
~.}
Notice that $r=0$ corresponds to chiral primary fields and $r=-1$ to
antichiral primary fields. The corresponding NS-sector characters
(for any $r\in \IZ$) read:
\eqn\aal{
\ch_d(h_{j,r},Q_{j,r};\tau,z)\bigg[ {0 \atop 0}\bigg]=
q^{\frac{-(j-1/2)^2+(j+r)^2}{k}} y^{\frac{2(j+r)}{k}}
\frac{1}{1+(-)^b y q^{\frac{1}{2}+r}}
\frac{\theta \big[ {0 \atop 0}\big](\tau,z)}{\eta(\tau)^3}
~.}
\item{($c$)} $Identity$ $representations$:
These representations are also degenerate and they have
quantum numbers $j=0$, $r\in \IZ$ with
\eqn\aama{
h_{r}=\frac{r^2}{k}-r-\frac{1}{2}~, ~ ~ Q_r=\frac{2r}{k}-1~, ~ ~ r<0
~,}
\eqn\aamb{
h_0=0~, ~ ~ Q_0=0~, ~ ~ r=0
~,}
\eqn\aamc{
h_r=\frac{r^2}{k}+r-\frac{1}{2}~, ~ ~ Q_r=\frac{2r}{k}+1~, ~ ~ r>0
~.}
The corresponding NS-sector characters (for any $r\in \IZ$) read:
\eqn\aan{\eqalign{
\ch_I(h_r,Q_r;\tau,z)\bigg[ {0 \atop 0} \bigg] =&
q^{-\frac{1}{4k}+\frac{r^2}{k}-r-\frac{1}{2}} y^{\frac{2r}{k}-1}
\cr
&\frac{1-q}{(1+(-)^b y^{-1}
q^{-\frac{1}{2}-r})(1+(-)^b y^{-1}q^{\frac{1}{2}-r})}
\frac{\theta \big[ {0 \atop 0}\big](\tau,z)}{\eta(\tau)^3}
~.}}

R-sector characters can be obtained by applying
the 1/2-spectral flow operation. To set the notation
straight we define the characters
\eqn\aao{\eqalign{
\ch_*(*;\tau,z)\bigg[ {0 \atop 0} \bigg]&=
\tr_{\rm NS}[q^{L_0-\frac{\hat c}{8}}
y^{J_0}]
\cr
\ch_*(*;\tau,z)\bigg[ {0 \atop 1} \bigg]&=\tr_{\rm \widetilde {NS}}
[(-)^{F} q^{L_0-\frac{\hat c}{8}}y^{J_0}]
\cr
\ch_*(*;\tau,z)\bigg[ {1 \atop 0} \bigg]&=\tr_{\rm R}
[q^{L_0-\frac{\hat c}{8}}y^{J_0}]
\cr
\ch_*(*;\tau,z)\bigg[ {1 \atop 1} \bigg]&=\tr_{\rm \widetilde R}
[(-)^{F} q^{L_0-\frac{\hat c}{8}}y^{J_0}]
~.}}
$*$ is an abbreviation for the specific representation and
$F$ denotes the total $fermion$ number.
As a simple illustration, for the continuous representations we obtain
the characters
\eqn\aap{\eqalign{
\ch_c(h_{j,m},Q_m;\tau,z)\bigg[{0\atop 0}\bigg]
&=q^{h_{j,m}-(\hat c-1)/8} y^{Q_m}
\frac{\theta \big[ {0 \atop 0}\big] (\tau,z)}{\eta(\tau)^3}
~,
\cr
\ch_c(h_{j,m},Q_m;\tau,z)\bigg[{0\atop 1}\bigg]
&= q^{h_{j,m}-(\hat c-1)/8} y^{Q_m}
\frac{\theta \big[ {0 \atop 1}\big] (\tau,z)}{\eta(\tau)^3}~,
\cr
\ch_c(h_{j,m+1/2},Q_{m+1/2};\tau,z)\bigg[{1\atop 0}\bigg]
&=q^{h_{j,m+\frac{1}{2}}-(\hat c-1)/8} y^{Q_{m+\frac{1}{2}}}
\frac{\theta \big[ {1 \atop 0}\big] (\tau,z)}{\eta(\tau)^3}
~,
\cr
\ch_c(h_{j,m+1/2},Q_{m+1/2};\tau,z)\bigg[{1\atop 1}\bigg]
&= q^{h_{j,m+\frac{1}{2}}-(\hat c-1)/8} y^{Q_{m+\frac{1}{2}}}
\frac{\theta \big[ {1 \atop 1}\big] (\tau,z)}{\eta(\tau)^3}
~.}}

The standard $\NN=2$ characters presented above
generate a continuous spectrum of $U(1)_R$ charges
under the modular transformation $\SS:\tau \rightarrow -\frac{1}{\tau}$.
This feature spoils the requirement
of charge integrality imposed by the
type II GSO projection. Hence, it is desirable to construct
a different set of ``extended'' characters that possess integral
$U(1)_R$ charges and at the same time form a closed set under
modular transformations. Such characters have been defined
in \EguchiIK\ for the cases with rational central charge
by taking appropriate sums over integer spectral flows of the
standard characters. Adapting the definition of \EguchiIK\
to the present situation of $k=1$ gives the extended characters
\eqn\aaqa{
\chi_c(s,m+\frac{a}{2};\tau,z)\bigg[ {a\atop b}\bigg]=\sum_{n \in \IZ}
\ch_c(h_{\frac{1}{2}+is,m+\frac{a}{2}+n},Q_{m+\frac{a}{2}+n};\tau,z)
\bigg[ {a\atop b}\bigg] ~, ~ ~ m=0~, \frac{1}{2}
~,}
\eqn\aaqb{
\chi_d(j,\frac{a}{2};\tau,z)\bigg[ {a\atop b}\bigg]=
\sum_{n \in \IZ} \ch_d(h_{j,\frac{a}{2}+n},Q_{j,\frac{a}{2}+n};\tau,z)
\bigg[ {a\atop b}\bigg] ~, ~ ~ j=\frac{\ell}{2}~, ~ \ell=1,2
~,}
\eqn\aaqc{
\chi_I(\tau,z)\bigg[ {a\atop b}\bigg]=
\sum_{n \in \IZ} \ch_I(h_{\frac{a}{2}+n},Q_{\frac{a}{2}+n};\tau,z)
\bigg[ {a\atop b}\bigg]
~.}
The $\SS$-modular transformation properties of these characters
are summarized in appendix A along with a useful set of character
identities. The torus partition function receives contributions from
the continuous and discrete representations only (see below). The
identity characters appear in the open string spectrum of
a special class of cigar D-branes.

\subsec{Type $0$ and type II non-critical superstring theory on $\IR^{3,1}\times SL(2)/U(1)$}

Type $0$ and type II non-critical superstring theory on
$\IR^{3,1} \times SL(2)_1/U(1)$ has been examined
previously in \refs{\MizoguchiKK,\EguchiTC,\MurthyES}.
Valuable information about the spectrum of these
theories can be obtained by analyzing the torus
partition function. This is also useful for implementing
appropriate constraints on the boundary states of the theory
later on. In general, the one-loop partition sum contains a volume-diverging
contribution from continuous representations and a finite contribution
from discrete representations. Both contributions can be
obtained using recent results on the torus partition function
of the bosonic and supersymmetric $SL(2)/U(1)$ coset in
\refs{\HananyEV,\EguchiYI,\IsraelIR,\FotopoulosUT,\EguchiIKK}. Here, we present
the resulting expressions for $k=1$ and summarize the basic features of the type II spectrum.
Earlier results on the continuous part of the type 0 and type II partition function
of the non-critical superstring $\IR^{3,1} \times SL(2)_1/U(1)$
have appeared in \refs{\MizoguchiKK,\EguchiTC,\MurthyES}.
Further details about the type II GSO projection
appear in appendix B.

The one-loop partition sum of the type $0$ theories
can be obtained by imposing a diagonal GSO projection
of the form
\eqn\baa{\eqalign{
{\rm 0A} ~ : ~ (-)^{J_{\rm GSO}}&=(-)^{\bar J_{\rm GSO}}~,
{\rm ~ ~ ~ ~ in ~ the ~ NS-sector}~,
\cr
(-)^{J_{\rm GSO}}&=(-)^{\bar J_{\rm GSO}+1}~, {\rm ~ in ~ the ~R-sector}~,
\cr
{\rm 0B} ~ : ~ (-)^{J_{\rm GSO}}&=(-)^{\bar J_{\rm GSO}}
~,}}
and the same fermion boundary conditions on the
left- and right-moving fermions.
The precise definitions of $J_{\rm GSO}$ and $\bar J_{\rm GSO}$ appear in
appendix B and include a sum on the fermion number of the flat
$\IR^{3,1}$ conformal field theory and the $U(1)_R$ charge of the supercoset.
The resulting one-loop partition sum takes the form
\eqn\bab{\eqalign{
Z_{\rm 0A/B}(\tau,\bar \tau)&=
\frac{1}{2}\sum_{a,b=0,1}\sum_{w\in \IZ_{2}} (-)^{\eta ab}\bigg\{
\int_0^{\infty} ds \sqrt{2} \rho(s,w,a;\epsilon)
\cr
&\chi_c\bigg(s,\frac{w+a}{2};\tau,0\bigg)\bigg[ {a \atop b}\bigg]
\chi_c\bigg(s,\frac{w+a}{2};\bar \tau,0\bigg)\bigg[ {a \atop b}\bigg]+
\cr
&+\frac{1}{2}
\chi_d\bigg(\frac{w}{2},\frac{a}{2};\tau,0\bigg)\bigg[ {a \atop b}\bigg]
\chi_d\bigg(\frac{w}{2},\frac{a}{2};\bar \tau,0\bigg)\bigg[ {a \atop b}\bigg]\bigg\}
\frac{\big|\theta\big[ { a \atop b }\big] \big|^2}{(8\pi^2 \tau_2)^2|\eta|^6}
~,}}
with spectral density
\eqn\bac{
\rho(s,w,a;\epsilon)=\frac{1}{\pi}\log\epsilon+\frac{1}{4\pi i}
\frac{d}{ds}\log\bigg\{
\frac{\Gamma(\frac{1}{2}-is+\frac{a+w}{2})\Gamma(\frac{1}{2}-is-\frac{a+w}{2})}
{\Gamma(\frac{1}{2}+is+\frac{a+w}{2})\Gamma(\frac{1}{2}+is-\frac{a+w}{2})}
\bigg\}
~.}
In this expression $\epsilon$ denotes the IR cutoff that
regularizes the infinite volume divergence
of the cigar CFT. $\eta=0/1$ corresponds to the type 0B/0A theory.

One can easily check that the volume
diverging piece of this partition sum is identical to the one
appearing in eq.\ (B.10) of \MurthyES. The extra discrete piece is
a by-product of the analysis appearing in refs.\
\refs{\HananyEV,\EguchiYI,\IsraelIR,\EguchiIKK}.
In our case ($k=1$), there are no discrete characters with half-integer $j$
inside the interval $\JJ:=(\frac{1}{2},\frac{k+1}{2}=1)$ and the
only discrete characters appearing in \bab\ are those lying on
the boundaries of $\JJ$. This extra contribution arises by defining
the integral over the continuous parameter $s$
with a principal value prescription that singles out
a pole at $s=0$ (for a nice exposition of the relevant details see
\IsraelIR).

To obtain the one-loop partition sum of the type II theory one
should perform a two-step procedure:
\item{($i$)} Impose the condition of integral $U(1)_R$ charges.
This condition is necessary for a well-defined
chiral GSO projection in step ($ii$) below. In the torus partition sum
\bab\ this integrality condition is automatic. Indeed, the characters
appearing in the type 0A/B partition sum have integral coset $U(1)_R$
charges in the NS-sector
\eqn\bad{
Q=2 \frac{w}{2} = w \in \IZ_2
~}
and the total fermion number is always an integer (see appendix B for further details).
\item{($ii$)} Perform the chiral GSO projection.
On the level of vertex operators this projection requires mutual locality
with respect to the spacetime supercharges of the theory and, similar
to the ten-dimensional critical case, it leads ultimately to a type IIA or type IIB
theory. In the non-critical case this prescription
has a peculiar feature (this point was emphasized in \MurthyES).
It leads to a non-trivial coupling
of the spin of the particles with their momentum around the angular
direction of the cigar and gives a spectrum that does not
have a natural spacetime interpretation as particles
propagating in six-dimensional curved spacetime. Instead,
the theory has a natural holographic interpretation as a non-gravitational
theory living in four dimensions.

\noindent Implementing the above procedure yields the following one-loop partition sum
\eqn\appggd{\eqalign{
Z_{\rm II}(\tau,\bar \tau)=&
\frac{1}{4}\sum_{a,\bar a,b,\bar b=0,1}\sum_{w \in \IZ_{2}}
(-)^{\eta ab+a+\bar a+(w+1)(b+\bar b)}\bigg\{
\int_0^{\infty} ds \sqrt{2} \rho(s,w;a,\bar a;\epsilon)
\cr
&\chi_c\bigg(s,\frac{w+a}{2};\tau,0\bigg)\bigg[ {a \atop b}\bigg]
\chi_c\bigg(s,\frac{w+\bar a}{2};\bar \tau,0\bigg)\bigg[ {\bar a \atop \bar b}\bigg]+
\cr
&+\frac{1}{2}
\chi_d\bigg(\frac{w}{2},\frac{a}{2};\tau,0\bigg)\bigg[ {a \atop b}\bigg]
\chi_d\bigg(\frac{w}{2},\frac{\bar a}{2};\bar \tau,0\bigg)\bigg[ {\bar a \atop \bar b}\bigg]\bigg\}
\frac{1}{(8\pi^2 \tau_2)^2 \eta^2 \bar \eta^2}\frac{\theta\big[ {a \atop b }\big]}
{\eta} \frac{\theta\big[ {\bar a \atop \bar b }\big]}{\bar \eta}
~,}}
where
\eqn\bae{
\rho(s,w;a,\bar a;\epsilon)=\frac{1}{\pi}\log\epsilon+\frac{1}{4\pi i}
\frac{d}{ds}\log\bigg\{
\frac{\Gamma(\frac{1}{2}-is+\frac{a+w}{2})\Gamma(\frac{1}{2}-is-\frac{\bar a+w}{2})}
{\Gamma(\frac{1}{2}+is+\frac{a+w}{2})\Gamma(\frac{1}{2}+is-\frac{\bar a+w}{2})}
\bigg\}
~.}
Again, one can check that the volume-diverging piece of this
partition sum is identical to the one appearing in \MizoguchiKK\
or \MurthyES\ (see eq.\ (B.13) of the latter paper).
By supersymmetry, we expect \appggd\ to be zero because of
the exact cancellation between bosons and fermions.
Indeed, we can check this explicitly for the continuous contributions
by writing everything in terms of the character combinations
\eqn\lambdaone{\eqalign{
\Lambda_{1}(s;\tau)&=
\Big(\chi_c(s,0;\tau,0)\bigg[ {0 \atop 0}\bigg]
\frac{\theta \big[{0 \atop 0} \big](\tau,0)}{\eta(\tau)^3}-
\chi_c(s,0;\tau,0)\bigg[ {0 \atop 1}\bigg]
\frac{\theta \big[{0 \atop 1} \big](\tau,0)}{\eta(\tau)^3}\Big)
\cr
&-\Big(\chi_c(s,\frac{1}{2};\tau,0)\bigg[ {1 \atop 0}\bigg]
\frac{\theta \big[{1 \atop 0} \big](\tau,0)}{\eta(\tau)^3}-
\chi_c(s,\frac{1}{2};\tau,0)\bigg[ {1 \atop 1}\bigg]
\frac{\theta \big[{1 \atop 1} \big](\tau,0)}{\eta(\tau)^3}\Big)
~,}}
\eqn\lambdamone{\eqalign{
\Lambda_{-1}(s;\tau)&=
\Big(\chi_c(s,\frac{1}{2};\tau,0)\bigg[ {0 \atop 0}\bigg]
\frac{\theta \big[{0 \atop 0} \big](\tau,0)}{\eta(\tau)^3}+
\chi_c(s,\frac{1}{2};\tau,0)\bigg[ {0 \atop 1}\bigg]
\frac{\theta \big[{0 \atop 1} \big](\tau,0)}{\eta(\tau)^3}\Big)
\cr
&-\Big(\chi_c(s,0;\tau,0)\bigg[ {1 \atop 0}\bigg]
\frac{\theta \big[{1 \atop 0} \big](\tau,0)}{\eta(\tau)^3}+
\chi_c(s,0;\tau,0)\bigg[ {1 \atop 1}\bigg]
\frac{\theta \big[{1 \atop 1} \big](\tau,0)}{\eta(\tau)^3}\Big)
~.}}
These combinations are known to be zero identically \refs{\BilalUH,\BilalIA}.
To check the vanishing of the discrete contributions one has to use in addition
the results of appendix A.

\vskip 15pt
\noindent
{\it A few comments on the closed string spectrum}
\vskip 3pt

Closing this section we would like to make a few final remarks
on the closed string spectrum following from the torus partition
function \appggd. A summarizing list of (the bosonic part of) this
spectrum from the six-dimensional point of view appears in
Table 1 below.

\vskip 20pt
\begintable
{\bf Theory} || {\bf Sector} | {\bf Fields} \crthick
IIA and IIB    ||   $NS+NS+$    | $G_{\mu \nu}, ~ B_{\mu \nu}, ~  \phi$
\cr
 ~ || $NS-NS-$ | $T, ~ T'$
\cr
IIA || $R+R-$ | $A_1$
\cr
~ || $R-R+$ | $A'_1$
\cr
IIB || $R+R+$ | $C_0, ~ C_2^+$
\cr
~ || $R-R-$ | $C'_0,~ C_2^-$
 \endtable
\nobreak\bigskip
{\raggedright\it \vbox{
{\bf Table 1.}
{\it The bosonic spectrum of type IIA and type IIB non-critical
superstring theory in \iae. The plus or minus superscripts for the RR potentials
denote the self-dual or anti-selfdual part respectively. The subscript denotes
the rank of the corresponding field. The fermionic part of the spectrum (NS-R sectors)
follows trivially by supersymmetry.
} }}
\vskip 15pt

The majority of fields appearing in this table
are massive. For instance, all the fields appearing in the NS$+$NS$+$
sector are massive including the graviton. Massless
fields arise from (continuous or discrete) representations with $j=\frac{1}{2}$
in the NS$-$NS$-$ and R$+$R$+$ sectors
(for simplicity we discuss only the bosonic sector here -
the fermionic sector can be determined easily by supersymmetry).
More precisely, from the NS$-$NS$-$ sector we obtain
two massless complex tachyons $T$, $T'$. One of them has
winding number $|w|=1$ and momentum zero and the other has winding
number zero and momentum $|n|=1$.
Physical massless states in the RR sector are
(from the six-dimensional point of view) in
the ${\bf 2} \times {\bf 2} = [0] + [2]_+$
representation of the little group $SO(4)$ for the type IIB theory
and in the ${\bf 2} \times {\bf 2'} = [1]$ for the type IIA theory.
In the type IIB case they correspond to a scalar $C_0$ and a self-dual 2-form
$C_2^+$. In the type IIA case they correspond to a vector $A_1$.
In both cases, these fields reduce to two scalars
and one vector in four dimensions, as expected from the unique non-chiral
structure of four-dimensional $\NN=2$ supersymmetry.

\newsec{Boundary conformal field theory on $\IR^{3,1}\times SL(2)/U(1)$}

In superstring theory it is standard to impose boundary
conditions preserving at least $\NN=1$ superconformal invariance
on the boundary of the worldsheet. In the closed string channel
this implies boundary conditions of the form
\eqn\flatbc{\eqalign{
(L_n-\bar L_{-n} )|B\rangle &=0,
\cr
(G_r-i \eta \bar G_{-r}) |B\rangle &=0,}}
where $\eta=\pm 1$ denotes the spin structure of the fermionic generators.

In the flat $\IR^{3,1}$ part of our theory
these conditions can be satisfied in the standard way familiar
from ten-dimensional critical superstring theory
\refs{\DiVecchiaRH, \DiVecchiaFX, \GaberdielJR}. In later parts
of this paper we want to consider D-brane configurations that
realize a $(3+1)$-dimensional gauge theory.
Hence, we have to impose Neumann boundary conditions
in all four flat directions of $\IR^{3,1}\times SL(2)/U(1)$ and
the corresponding Ishibashi states will be characterized by
a vanishing momentum and the spin structure of the fermions.
These states will be denoted simply as
\eqn\flatishii{
|p_{\mu}=0;[{a \atop b}]\rrangle \equiv|[{a \atop b}]\rrangle_{\rm flat}
}
and they have a standard construction
as coherent states in the free supersymmetric $\IR^{3,1}$ conformal field theory.
In the covariant formalism, which is the formalism we are implicitly adopting,
one should include also the contribution of ghosts. The explicit
form of the ghost boundary states can be found in \DiVecchiaRH.
In \flatishii\ the label $a=0,1$ parametrizes a boundary state in
the NSNS and RR sectors respectively, while the second label
$b=0,1$ parametrizes the choice of spin structure $\eta$.
The corresponding cylinder amplitudes take the form
\eqn\flatishiover{
_{\rm flat}\bllangle  \bigg[{a' \atop b'}\bigg]\big|e^{-\pi T H^c_{\rm flat}}\big|
\bigg[{a \atop b}\bigg]\rrangle_{\rm flat} = (-)^a
\delta_{a,a'} \frac{\theta\big[ {a \atop b-b' }\big](iT,0)}{\eta^3(iT)}
~.}

In $SL(2)/U(1)$ we choose to impose a more symmetric
set of boundary conditions preserving $\NN=2$ superconformal invariance
on the boundary of the worldsheet.
These are the well-known boundary conditions \OoguriCK:
\eqn\appeea{
{\rm A-type} ~ : ~ (J_n-\bar J_{-n})|B\rangle =0~, ~ ~
(G_r^{\pm}-i\eta \bar G_{-r}^{\mp})|B\rangle =0~,
}
\eqn\appeeb{
{\rm B-type} ~ : ~ (J_n+\bar J_{-n})|B\rangle =0~, ~ ~
(G_r^{\pm}-i\eta \bar G_{-r}^{\pm})|B\rangle=0
~.}
The A-type boundary conditions are Neumann
in the angular direction of the cigar and
the B-type are Dirichlet.\foot{In more standard conventions
($c.f.$ \refs{\EguchiIK, \IsraelJT}) A-type boundary conditions
are always Dirichlet and B-type are Neumann. In this paper we use
the opposite convention associated with the right-moving
$\NN=2$ current $\bar J_{\NN=2}=\bar{\psi}^+\bar{\psi}^-+\frac{2}{k} \bar{\JJ}^3$.}
Corresponding Ishibashi states can be constructed based
on continuous or discrete representations.
These will be denoted as
$|X;s,m,\bar m; [{a \atop b}]\rrangle_{\rm cos}$
for the continuous representations
and $|X;j; [{a \atop b}]  \rrangle_{\rm cos}$
for the discrete. $X=A$, $B$ is an extra label specifying
the type of boundary condition and the parameters
$s, m, \bar m, j$ take the appropriate values dictated by the representations
appearing in the torus partition sum and the specific boundary conditions.
The corresponding cylinder amplitudes are
\eqn\cigarishiover{\eqalign{
_{\rm cos}\bllangle X; s, m, \bar m ; \bigg[{a \atop b} \bigg]
\big|e^{-\pi T H^c_{\rm coset}}
\big| X; s' , m', \bar  m'; \bigg[{a' \atop b'}\bigg] \brrangle_{\rm cos}
=&\delta_{a,a'} \delta(s-s') \delta_{m,m'}
\cr
&\chi_c(s,m; iT,0)\bigg[ {a\atop b'-b}\bigg]
~,
\cr
_{\rm cos}\bllangle X; j; \bigg[{a \atop b} \bigg]\big|e^{-\pi T H^c_{\rm coset}}
\big| X; j'; \bigg[{a' \atop b'}\bigg] \brrangle_{\rm cos}=&
\delta_{a,a'} \delta_{j,j'} \chi_d(j,\frac{a}{2}; iT,0)\bigg[ {a\atop b'-b}\bigg]
~.}}

The Ishibashi states of the full theory are tensor products of the $\IR^{3,1}$
Ishibashi states $|[{a \atop b}]\rrangle_{\rm flat}$
with A- or B-type Ishibashi states of the coset.
However, the generic tensor product is not an allowed Ishibashi state.
Only those states that couple to the closed string
modes appearing in the torus partition sum \appggd\ are allowed.
This implies a set of constraints.

First, we have a constraint on the
combination of spin structures. The same spin structure must appear
on the flat and coset components, $i.e.$ we should restrict to
boundary states of the form
\eqn\ishispina{
\big|X;s,m,\bar m; \bigg[{a \atop b}\bigg]\rrangle=
\big|\bigg[{a \atop b}\bigg]\rrangle_{\rm flat}
\otimes \big|X;s,m,\bar m; \bigg[{a \atop b}\bigg]\rrangle_{\rm cos}
}
and
\eqn\ishispinb{
\big|X;j; \bigg[{a \atop b}\bigg]\rrangle =
\big|\bigg[{a \atop b}\bigg]\rrangle_{\rm flat}
\otimes \big|X;j; \bigg[{a \atop b}\bigg]\rrangle_{\rm cos}
~.}
This can be rephrased
as the requirement to have a well-defined periodicity for
the total $\NN=1$ supercurrent
$G_{\rm total}=G_{\rm flat}+G_{\rm coset}^++G_{\rm coset}^-$.

A second set of constraints comes from GSO invariance.
For simplicity, let us consider here only the type IIB case.
By simple inspection of the torus partition sum \appggd, or
by explicitly checking how $(-)^{J_{\rm GSO}}$, $(-)^{\bar J_{\rm GSO}}$
act on the Ishibashi states and requiring
$(-)^{J_{\rm GSO}}=(-)^{\bar J_{\rm GSO}}=1$,
we find a set of GSO-allowed linear superpositions of Ishibashi states.
For example, the allowed NSNS continuous Ishibashi states are
\eqn\nsfullishi{\eqalign{
\big|A;s,0,0;+ \brrangle_{NS} &= \big|A;s,0,0; \bigg[{0 \atop 0}\bigg]\brrangle
- \big|A;s,0,0; \bigg[{0 \atop 1}\bigg]\brrangle ~,
\cr
\big|A;s,\frac{1}{2},\frac{1}{2};- \brrangle_{NS} &=
\big|A;s,\frac{1}{2},\frac{1}{2}; \bigg[{0 \atop 0}\bigg]\brrangle
+ \big|A;s,\frac{1}{2},\frac{1}{2}; \bigg[{0 \atop 1}\bigg]\brrangle
~.}}
Notice the correlation between the quantum numbers $m, \bar m$ and
the sign of total fermion chirality $(-)^{F_{\rm fermion}+a-1}$,
which appears as an extra index $\pm$ in
the Ishibashi state. The corresponding RR sector Ishibashi states
take the form
\eqn\rfullishi{\eqalign{
\big|A;s,0,0;+ \brrangle_{R} &= \big|A;s,0,0; \bigg[{1 \atop 0}\bigg]\brrangle
+ \big|A;s,0,0; \bigg[{1 \atop 1}\bigg]\brrangle ~,
\cr
\big|A;s,\frac{1}{2},\frac{1}{2};- \brrangle_{R} &=
\big|A;s,\frac{1}{2},\frac{1}{2}; \bigg[{1 \atop 0}\bigg]\brrangle
- \big|A;s,\frac{1}{2},\frac{1}{2}; \bigg[{1 \atop 1}\bigg]\brrangle
~.}}
The flip of sign conventions between the NSNS and RR sectors
is due to the superconformal ghost contribution to $(-)^{F_{\rm fermion}+a-1}$.
Similar expressions can be written for the A-type discrete
states and for the B-type NSNS Ishibashi states. The B-type
RR Ishibashi states have $(-)^{J_{\rm GSO}}=-(-)^{\bar J_{\rm GSO}}=1$
and they have to be excluded in type IIB string theory.
This point has important consequences for the BPS spectrum
of branes in this theory and we would like to explain it here in some detail.

Working in the covariant formalism we can write the full GSO charge
in the Ramond sector as
\eqn\bca{
J_{\rm GSO}=F_{\rm flat}+J_{\NN=2}-\frac{1}{2}
}
and this should be an even integer for GSO projected states.
The last term $-\frac{1}{2}$ comes from the superghost contribution.
$F_{\rm flat}$ denotes the flat space fermion number
\eqn\bcb{
F_{\rm flat}=s_0+s_1~, ~ ~ s_0, ~ s_1=\pm \frac{1}{2}
}
and $J_{\NN=2}$ is the $U(1)_R$ charge
\eqn\bcc{
J_{\NN=2}=2m_R+\frac{1}{2}
~.}
The half-integer $m_R$ is the R-sector ${\cal J}^3$ charge of $SL(2)/U(1)$.
For B-type boundary conditions the right-moving charges are related to the left ones
by the following equations
\eqn\bcd{
F_{\rm flat}=\bar F_{\rm flat}~, ~ ~ J_{\NN=2}=-\bar J_{\NN=2}
~.}
Hence,
\eqn\bce{
\bar J_{\rm GSO}=\bar F_{\rm flat}+\bar J_{\NN=2}-\frac{1}{2}=s_0+s_1-2m_R-1
}
and
\eqn\bcf{
(-)^{\bar J_{\rm GSO}}=(-)^{s_0+s_1-2m_R-1}=-(-)^{J_{\rm GSO}}
~}
as claimed above.

Implementing the full set of the above constraints we find the
allowed Ishibashi states
\item{$\bullet$} {\it A-type, continuous:}
\eqn\atypecontfull{\eqalign{
&|A;s,0,0;+\rrangle_{NS},\; |A;s,\frac{1}{2},\frac{1}{2};-\rrangle_{NS}~,
\cr
&|A;s,0,0;+\rrangle_{R},\; |A;s,\frac{1}{2},\frac{1}{2};-\rrangle_{R}, \;\;\;  s\in \IR_+
~,}}
\item{$\bullet$} {\it B-type, continuous:}
\eqn\btypecont{
|B;s,0,0;+\rrangle_{NS},\; |B;s,\frac{1}{2},-\frac{1}{2};-\rrangle_{NS}
~, ~ ~ s\in \IR_+
~.}

\noindent
Similar discrete A-type Ishibashi states exist, but they will not be mentioned
here explicitly, since they play no role in the boundary state
analysis of the next subsections.

In what follows we employ these results to formulate
and analyze the properties of D-branes in the four-dimensional
non-critical superstring theory under consideration.

\subsec{A-type boundary states}

In this subsection we formulate A-type boundary states
as appropriate linear combinations of the Ishibashi states presented above.
The coefficients can be determined by using previously obtained results
on the boundary states of the coset $SL(2)/U(1)$.
In some cases these coefficients follow directly from a generalized Cardy ansatz,
but there are also situations where one has to use slight variants that have been derived
by different methods. Here we discuss each case in detail and
explain any potential subtleties. At the end, we verify the Cardy consistency
conditions by a straightforward computation of the annulus amplitudes.

A generic A-type boundary state labelled by $\xi$
will be written in the NS and R-sectors as
\eqn\caa{
|A;\xi \rangle\rangle_{NS}=\int_0^{\infty} ds
~ \big( \Phi_{NS}(s,+;\xi) |A;s,0;+\rrangle_{NS}+
\Phi_{NS}(s,-;\xi) |A;s,\frac{1}{2};-\rrangle_{NS} \big)
~,}
\eqn\cab{
|A;\xi \rangle\rangle_{R}=\int_0^{\infty} ds
~ \big( \Phi_{R}(s,+;\xi) |A;s,0;+\rrangle_{R}+
\Phi_{R}(s,-;\xi) |A;s,\frac{1}{2};-\rrangle_{R} \big)
~.}
$\xi$ will be an index or a set of indices characterizing the $SL(2)/U(1)$ properties of
the brane. In principle, $\xi$ can be a label corresponding to continuous,
discrete or identity representations, but a more precise analysis reveals the
following possibilities.


\bigskip
\noindent
\mathboxit{\it Class ~ 1}
\smallskip
\noindent
Boundary states in this class are based on the $identity$ representation
and will be denoted as $|A\rangle_{NS}$ and $|A\rangle_R$. They can be obtained
from a direct application of the Cardy ansatz, which implies in our case
the following wavefunctions\foot{Here and below we do not include
a standard phase factor $\nu_k^{is}$, with $\nu_k=\frac{\Gamma(1-\frac{1}{k})}
{\Gamma(1+\frac{1}{k})}$, because it diverges for $k=1$. This factor does
not affect the computation of annulus amplitudes.}
\eqn\cac{
\Phi_{NS}(s,+;I)=\Phi_R(s,-;I)=\frac{1}{2}
\sqrt{S^c(s,0;\bigg[ {0 \atop 0} \bigg] | I;\bigg[ {0 \atop 0} \bigg])}
= \sinh(\pi s)
~,}
\eqn\cad{
\Phi_{NS}(s,-;I)=\Phi_R(s,+;I)=\frac{1}{2}
\sqrt{S^c(s,\frac{1}{2};\bigg[ {0 \atop 0} \bigg] | I;\bigg[ {0 \atop 0} \bigg])}
=\cosh(\pi s)
~.}
The corresponding reflection-invariant one-point functions on the disc are
\eqn\caca{\eqalign{
\langle \VV^{NS+NS+}_{\frac{1}{2}+is,m,\bar m}(p^{\mu})\rangle=
\langle \VV^{R-R-}_{\frac{1}{2}+is,m+\frac{1}{2},\bar m+\frac{1}{2}}(p^{\mu})\rangle=&
\delta^{(4)}(p^{\mu})\delta_{m,\bar m}
\frac{1}{2}\frac{\Gamma (\frac{1}{2}+is+m)\Gamma(\frac{1}{2}+is-m)}
{\Gamma(1+2is)\Gamma(2is)}\cr
\sim&\delta^{(4)}(p^{\mu})\delta_{m,\bar m} \sinh(\pi s)~, ~ ~ m\in \IZ
~,}}
\eqn\cacb{\eqalign{
\langle \VV^{NS-NS-}_{\frac{1}{2}+is,m,\bar m}(p^{\mu})\rangle=
\langle \VV^{R+R+}_{\frac{1}{2}+is,m+\frac{1}{2},\bar m+\frac{1}{2}}(p^{\mu})\rangle=&
\delta^{(4)}(p^{\mu})\delta_{m,\bar m}
\frac{1}{2}\frac{\Gamma (\frac{1}{2}+is+m)\Gamma(\frac{1}{2}+is-m)}
{\Gamma(1+2is)\Gamma(2is)}
\cr
\sim & \delta^{(4)}(p^{\mu})\delta_{m,\bar m} \cosh(\pi s)~, ~ ~
m\in \IZ +\frac{1}{2}
~.}}
The similarity symbol $\sim$ denotes equality up to a phase and
$p^{\mu}$ is the four-dimensional Minkowski space momentum.

These boundary states correspond to D3-branes
and can be thought of as the analogs of
the Liouville theory ZZ-branes. Geometrically, they are localized
near the tip of the cigar (see fig.\ 2) with a smooth profile along
the radial direction. In general, there are two
clear signals of the localization of this class of branes
near the tip: the vanishing of some of the continuous wavefunctions for
zero radial momentum $s$ and the presence of discrete couplings.
The first property is apparent in \cac, but the second is not as a
consequence of the very special features of the $k=1$ case.

\bigskip
{\vbox{{\epsfxsize=35mm
\nobreak
\centerline{\epsfbox{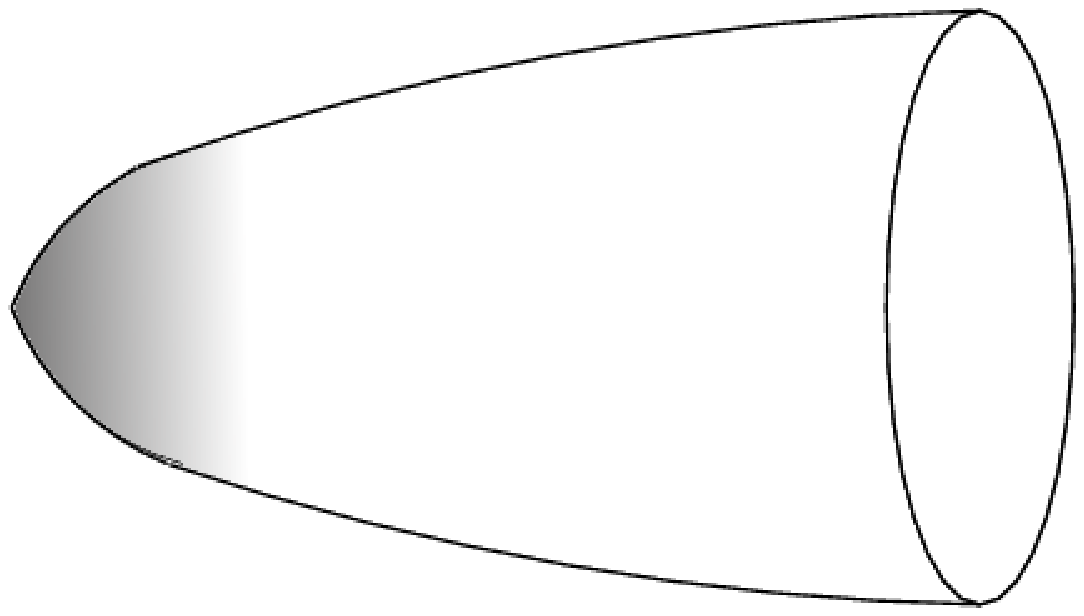}}
\nobreak\bigskip
{\raggedright\it \vbox{
{\bf Figure 2.}
{\it D3-branes have a smooth profile in the radial direction of the cigar supported
near the tip.
} }}}}
\bigskip}

\noindent
\mathboxit{\it Class ~ 2}
\smallskip
\noindent
In this class we consider boundary states based on the $continuous$ representations.
They will be denoted as $|A;s,m\rangle_{NS}$ and $|A;s,m\rangle_{R}$, with
parameters $s \in \IR_{\geq 0}$ and $m=0,\frac{1}{2}$.
On $SL(2)_k/U(1)$ (for even levels $k$) these branes were first
formulated in \FotopoulosUT. There it was argued that they correspond to
D2-branes partially or totally covering the cigar, where
$s$ is a modulus parametrizing the closest distance between the brane
and the tip (for the semiclassical analysis
of these branes see \FotopoulosVC).

The precise form of their wavefunctions (for generic integer level
$k$) can be determined in the following way. Starting from the
T-dual trumpet geometry, which strictly speaking is described by
the $\NN=2$ Liouville theory, we can formulate B-type D1-branes
which extend in the radial direction. The expressions and
consistency of the wavefunctions of the corresponding boundary
states has been determined by direct computation with modular and
conformal bootstrap methods in \HosomichiPH.\foot{Because of
different conventions these are A-type boundary states in
\HosomichiPH\ - see eq.\ (3.21) in that paper.} After a T-duality transformation
the resulting expressions for the A-type class 2 cigar boundary states at $k=1$ are:
\eqn\caf{\eqalign{ \Phi_{NS}(s',+;s,m)&=(-1)^{2m}\Phi_R(s',-;s,m)=
\frac{e^{4\pi i ss'}+e^{-4\pi i ss'}}{4\sinh(\pi s')}~, \cr
\Phi_{NS}(s',-;s,m)&=(-1)^{2m}\Phi_R(s',+;s,m)=(-1)^{2m} \frac{
e^{4\pi i ss'}+e^{-4\pi i ss'}}{4\cosh(\pi s')} ~,}} and the
corresponding reflection-invariant one-point functions on the disc
\eqn\cafa{\eqalign{ &\langle \VV^{NS+NS+}_{\frac{1}{2}+is',m',\bar
m'}(p^{\mu})\rangle_{s,m}= (-)^{2m}\langle
\VV^{R-R-}_{\frac{1}{2}+is',m'+\frac{1}{2}, \bar
m'+\frac{1}{2}}(p^{\mu})\rangle_{s,m}=\cr
&\delta^{(4)}(p^{\mu})\delta_{m',\bar m'} \frac{\Gamma
(1-2is')\Gamma(-2is')}
{\Gamma(\frac{1}{2}-is'+m')\Gamma(\frac{1}{2}-is'-m')} \cos(4\pi
ss')~, ~ ~ m'\in \IZ ~,}} \eqn\cafb{\eqalign{ &\langle
\VV^{NS-NS-}_{\frac{1}{2}+is',m',\bar m'}(p^{\mu})\rangle_{s,m}=
(-)^{2m}\langle \VV^{R+R+}_{\frac{1}{2}+is',m'+\frac{1}{2}, \bar
m'+\frac{1}{2}}(p^{\mu})\rangle_{s,m}=\cr
&(-)^{2m}\delta^{(4)}(p^{\mu})\delta_{m',\bar m'} \frac{\Gamma
(1-2is')\Gamma(-2is')}
{\Gamma(\frac{1}{2}-is'+m')\Gamma(\frac{1}{2}-is'-m')} \cos(4\pi
ss')~, ~ ~ m'\in \IZ +\frac{1}{2} .}} $s$ is a non-negative real
number and $m=0,\frac{1}{2}$.
We should emphasize that these
boundary states are automatically consistent because they have
been derived by T-duality from consistent branes of the $\NN=2$
Liouville theory.

Later in this section we will see that these
branes contain open string states with
both integer and half-integer momenta. This implies that
the class 2 boundary states appearing in \caf\ describe
a superposition of branes with a $U(2)$ gauge symmetry
broken down to $U(1) \times U(1)$ by the presence of a Wilson line.
An alternative but equivalent picture of the same effect is provided by the
corresponding D1-brane on the T-dual trumpet.
This brane has two branches as well (see fig.\ 3)
and the open strings have integer or half-integer winding numbers
depending on whether they stretch between the same or different branches.
The angular separation of the two branches by an angle $\Delta \theta=\pi$
translates after T-duality to a non-trivial Wilson line between the two
``sheets'' of the cigar D2-brane.

\bigskip
{\vbox{{\epsfxsize=35mm
\nobreak
\centerline{\epsfbox{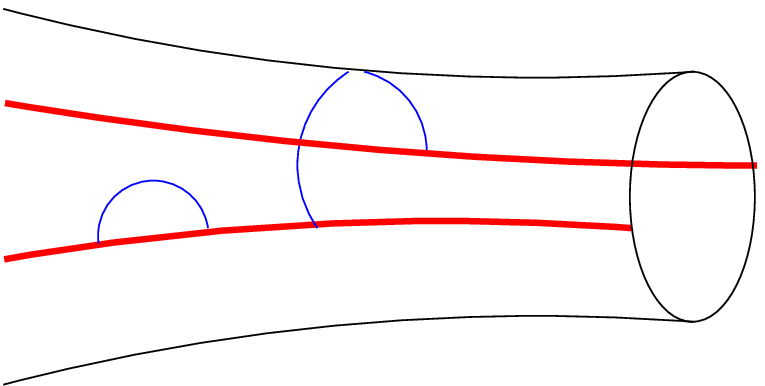}}
\nobreak\bigskip
{\raggedright\it \vbox{
{\bf Figure 3.}
{\it A D1-brane with two branches on the T-dual trumpet geometry. Open strings
stretching on the same branch have integer windings whereas
open strings stretching between different branches have half-integer windings.
This configuration maps to a double-sheeted D2-brane on the cigar.
} }}}}
\bigskip}

One may be tempted to associate the
two exponentials $e^{\pm 4\pi i ss'}$ in the wavefunctions \caf\ to
the two more fundamental sheets that have
different orientations. If we do that, we find that the spectrum of the resulting
branes contains again both integer and half-integer momenta.
This is not what we expect from decomposed one-sheeted D2-branes.
Trying to further decompose these boundary states by separating
different exponential contributions in the wavefunctions
leads to boundary states that violate the Cardy consistency conditions
with the class 1 brane. Hence, such decompositions do not appear to be
admissible and they will not be discussed further in this paper.

\bigskip
\noindent
\mathboxit{\it Class ~ 3}
\smallskip
\noindent
According to the general discussion of D-branes in $SL(2)/U(1)$,
this class should contain boundary states with open strings in the
$discrete$ representations. In the present case, there are only two discrete
representations (with $j=\frac{1}{2},1$) and they are both closely related to
the continuous representation with $s=0$. Hence, the application of the
modular bootstrap does not lead to a genuinely new class of branes. It
simply reproduces a class 2 boundary state with $s=0$.

A different class of D2-branes (dubbed D2 cut branes in \IsraelFN)
has been formulated for generic levels $k$
in \refs{\RibaultSS, \IsraelJT}. In general, these branes have
negative multiplicities in the open string channel and do not satisfy the Cardy
consistency conditions. Recently, it was argued in \IsraelFN\ that
this problem does not exist for integer levels $k$, because the dangerous
discrete couplings in the closed string channel disappear.
These branes are labeled by a single parameter
$\sigma=\frac{\pi (2J-1)}{k}$, with $2J\in \IN$ and $\frac{1}{2}<J<\frac{k+1}{2}$.
For $k=1$ there are no $J$'s in this range.
It is interesting to notice, however, that the special case
$\sigma=\frac{\pi}{2}$, or $J=\frac{3}{4}$, reproduces
the $s=0$ class 2 boundary state of the previous paragraph.

\subsec{B-type boundary states}

The analysis of B-type boundary states is technically similar
to that of the A-type boundary states appearing above and we will not
repeat it here. There are a few differences, however, which
should be pointed out. First, as we mentioned earlier, the Ramond
part of the B-type Ishibashi states is projected out by the GSO projection.\foot{Recall
that we are considering the type IIB superstring and D-branes that have
Neumann boundary conditions in all four flat directions.}
Thus, all the B-type boundary states (with Neumann boundary
conditions in the flat directions) will be non-BPS.
A second important point is the absence
of consistent B-type class 1 boundary states. This was argued for generic
levels $k$ (integers included) in \HosomichiPH.
Consequently, one is left with a set of class 2 boundary
states in the NSNS sector only, which can be formulated as above
(with a few appropriate modifications in the wavefunctions).

\subsec{Cylinder amplitudes}

In this subsection we compute the cylinder/annulus amplitudes of the
above class 1 and class 2 boundary states.
The modular transformation of these amplitudes
from the closed string channel (parameter $T$) to the open (parameter $t=1/T$)
yields the explicit form of the spectral densities and the degeneracies of the open
strings stretching between the various branes. We omit a detailed analysis
of A-B and B-B overlaps, because they involve non-supersymmetric
D-brane configurations that lie outside the immediate scope of this paper.

\vskip 15pt
\noindent
\mathboxit{\it class ~ 1 - class ~1}
\smallskip
\noindent
By straightforward computation we find the following annulus amplitudes
between class 1 boundary states:\foot{In the rhs of the annulus amplitudes
that appear in the ensuing, a factor of $\frac{1}{t}$ is omitted for simplicity.}
\eqn\daa{
_{NS}\langle A|e^{-\pi T H^c}|A\rangle_{NS}=
\frac{1}{2}\bigg(\chi_I(it)\bigg[ {0 \atop 0}\bigg]
\frac{\theta [{0 \atop 0} ](it)}{\eta(it)^3} - \chi_I(it)\bigg[ {1 \atop 0}\bigg]
\frac{\theta [{1 \atop 0} ](it)}{\eta(it)^3}\bigg)
~,}
\eqn\dab{
_{R}\langle A|e^{-\pi T H^c}|A\rangle_{R}=
\frac{1}{2}\bigg(-\chi_I(it)\bigg[ {0 \atop 1}\bigg]
\frac{\theta [{0 \atop 1} ](it)}{\eta(it)^3} + \chi_I(it)\bigg[ {1 \atop 1}\bigg]
\frac{\theta [{1 \atop 1} ](it)}{\eta(it)^3}\bigg)
~.}
The boundary state describing a BPS D3-brane
is $|A\rangle = |A\rangle_{NS} + |A\rangle_R$,
whereas that describing a D3-antibrane is
$|\overline{A}\rangle = |A\rangle_{NS} - |A\rangle_R$.
The self-overlaps of these boundary states are the same
\eqn\dac{\eqalign{
\langle A |e^{-\pi T H^c}| A \rangle =
\langle \overline{A} |e^{-\pi T H^c}| \overline{A} \rangle
&=
\frac{1}{2}\bigg(\chi_I(it)\bigg[ {0 \atop 0}\bigg]
\frac{\theta [{0 \atop 0} ](it)}{\eta(it)^3} - \chi_I(it)\bigg[ {1 \atop 0}\bigg]
\frac{\theta [{1 \atop 0} ](it)}{\eta(it)^3}
\cr
& -\chi_I(it)\bigg[ {0 \atop 1}\bigg]
\frac{\theta [{0 \atop 1} ](it)}{\eta(it)^3} + \chi_I(it)\bigg[ {1 \atop 1}\bigg]
\frac{\theta [{1 \atop 1} ](it)}{\eta(it)^3}\bigg)
~.}}
As expected by supersymmetry both of them are vanishing.
This can be demonstrated most easily in the closed
string channel with the use of the vanishing character combinations
 $\Lambda_{\pm 1}(s;\tau)$ in \lambdaone, \lambdamone.

\vskip 15pt
\noindent
\mathboxit{\it class ~2 - class ~2}
\smallskip
\noindent
The class 2 boundary states $|A;s,m\rangle_{NS/R}$, defined in
\caf, exhibit the following amplitudes. In the NS-sector
\eqn\daf{\eqalign{
&_{NS}\langle A;s_1,m_1|e^{-\pi T H^c}|A;s_2,m_2 \rangle_{NS}=\cr
&=\int_0^{\infty} ds \sum_{m\in \IZ_2}\bigg(
\Big(\rho_1(s;s_1|s_2) +(-1)^{2m_1+2m_2+m} \rho_2(s;s_1|s_2) \Big)
\chi_c(s,\frac{m}{2};it)\bigg[ {0 \atop 0}\bigg]
\frac{\theta [{0 \atop 0} ](it)}{\eta(it)^3}\cr
&-\Big(\rho_1(s;s_1|s_2) -  (-1)^{2m_1+2m_2+m} \rho_2(s;s_1|s_2) \Big)
\chi_c(s,\frac{m}{2};it)\bigg[ {1 \atop 0}\bigg]
\frac{\theta [{1 \atop 0} ](it)}{\eta(it)^3} \bigg)
~,}}
with spectral densities
\eqn\dag{
\rho_1(s;s_1|s_2)=2\;\int_0^{\infty} ds'
\frac{\cos(4\pi s' s_1)\cos(4\pi s' s_2)\cos(4\pi s s')}
{\sinh (2\pi s') \tanh (\pi s')}
~,}
and
\eqn\dai{
\rho_2(s;s_1|s_2)=2\;\int_0^{\infty} ds'
\frac{\cos(4\pi s' s_1)\cos(4\pi s' s_2)\cos(4\pi s s')}
{\sinh (2\pi s') \coth (\pi s')}
~.}
Similarly, in the R-sector
\eqn\daff{\eqalign{
&_{R}\langle A;s_1,m_1|e^{-\pi T H^c}|A;s_2,m_2 \rangle_{R}=\cr
&=-\;\int_0^{\infty} ds \sum_{m\in \IZ_2}\bigg(
\Big((-1)^{2m_1+2m_2+m} \rho_1(s;s_1|s_2) +\rho_2(s;s_1|s_2) \Big)
\chi_c(s,\frac{m}{2};it)\bigg[ {0 \atop 1}\bigg]
\frac{\theta [{0 \atop 1} ](it)}{\eta(it)^3}\cr
&+\Big((-1)^{2m_1+2m_2+m}\rho_1(s;s_1|s_2) -  \rho_2(s;s_1|s_2) \Big)
\chi_c(s,\frac{m}{2};it)\bigg[ {1 \atop 1}\bigg]
\frac{\theta [{1 \atop 1} ](it)}{\eta(it)^3} \bigg)
~.}}

The total densities appearing in front of the continuous characters
in the above amplitudes are $\rho_1(s;s_1|s_2) \pm \rho_2(s;s_1|s_2)$
depending on the precise values of $m_1, m_2$ and $m$.
The spectral density $\rho_1(s;s_1|s_2)$ has an infrared divergence
at $s'=0$ associated to the infinite volume of the non-compact cigar geometry.
As usual, this divergence can be regulated by subtracting the
amplitude of a reference boundary state labeled by $s^*$.
We will not specify a particular reference brane here.

In quantum theories with reflecting potentials there is a
general relation between the density of continuous states
and the appropriate reflection amplitudes (for a review of this argument
see \PonsotGT). We can verify this relation explicitly in our case.
Indeed, we obtain
\eqn\densityinteger{
\rho_1(s;s_1|s_2) + \rho_2(s;s_1|s_2)\Big|_{\rm rel} = \frac{1}{2\pi i}
\frac{\partial}{\partial s}
\Big (\log\frac{R(s,\frac{1}{2}|\pi (s_1+ s_2))}{R(s,\frac{1}{2}|2\pi  s^*)} +
\log\frac{R(s,\frac{1}{2}|\pi (s_1- s_2))}{R(s,\frac{1}{2}| 0)}\Big)
~,}
\eqn\densityhalfinteger{
\rho_1(s;s_1|s_2) - \rho_2(s;s_1|s_2)\Big|_{\rm rel} = \frac{1}{2 \pi i}
\frac{\partial}{\partial s} \Big (\log\frac{R(s,0|\pi (s_1- s_2))}{R(s,0|0)} +
\log\frac{R(s,0|\pi (s_1+ s_2))}{R(s,0| 2\pi s^*)}\Big)
~,}
with reflection amplitudes
\eqn\reflectionone{
R(s,0 |r) = \frac{\Gamma_1^2(\frac{1}{2}-is)
\Gamma_1(2 i s + 1) S_1^{(0)}(s+
\frac{r}{\pi})}{\Gamma_1^2(\frac{3}{2}+is) \Gamma_1(-2is+1)
S_1^{(0)}(-s+ \frac{r}{\pi})}
}
for integer momenta, and
\eqn\reflectiontwo{
R(s,\frac{1}{2} |r) =  \frac{\Gamma_1^2(\frac{1}{2}-is)
\Gamma_1(2 i s + 1) S_1^{(1)}(s+
\frac{r}{\pi})}{\Gamma_1^2(\frac{3}{2}+is) \Gamma_1(-2is+1)
S_1^{(1)}(-s+ \frac{r}{\pi})}
~} for half-integer momenta.
The q-gamma functions $S_1^{(0)}(x)$ and $S_1^{(1)}(x)$
are defined as
\eqn\qgamma{
\log S_k^{(0)}(x) = i \int_0^{\infty}\frac{dt}{t}
\left( \frac{\sin \frac{2tx}{k}}{2 \sinh \frac{t}{k} \sinh t}-\frac{x}{t}
\right)
~,}
\eqn\qgammanext{
\log S_k^{(1)}(x) = i \int_0^{\infty}\frac{dt}{t}
\left( \frac{\cosh t \sin \frac{2tx}{k}}{2 \sinh \frac{t}{k} \sinh
t}-\frac{x}{t} \right)
~.}
The generalized gamma functions $\Gamma_k$ can be found, for example in \RibaultSS.
We do not present the explicit form of these functions here since they
cancel out in the full eqs.\ \densityinteger\ and \densityhalfinteger\
for the relative densities. Similar expressions for the spectral densities have been
found in \RibaultSS\ and \IsraelJT.

At this point we would like to make two comments.
First, for a single brane, $i.e.$ for an amplitude with $s_1=s_2=s$ and $m_1=m_2$,
the spectral density of modes with momentum $m$
appears as a function of the reflection amplitude
with quantum number $m+\frac{1}{2}$ mod $1$.
For instance, the density of open string modes
with integer momentum $m$ in the NS-sector
is $\rho_1(s;s_1|s_1) + \rho_2(s;s_1|s_1)\Big|_{\rm rel}$.
In \densityinteger\ we see that the corresponding
reflection amplitude is $ R(s,\frac{1}{2} |2\pi s_1)$.
It would be interesting to understand this feature better.
Secondly, with the current normalization of the class 2 branes \caf\
the expressions \densityinteger\ and \densityhalfinteger\
agree with the general formula
$\rho(s) = \frac{1}{2\pi i} \frac{\partial}{\partial s} \log\frac{R(s)}{R^*(s)}$.
The current normalization of the class 2 branes can also be
fixed independently by requiring that the class 1-class 1 and
class 1-class 2 overlaps give the expected multiplicity
of massless open string modes. Further arguments
in favor of this normalization and the associated multiplicities
will be given in section 5.

BPS boundary states can be formulated as before. They are given
by the linear combinations
\eqn\dafa{\eqalign{
|A;s,m\rangle &= |A;s,m\rangle_{NS} + |A;s,m\rangle_R
\cr
|\overline{A;s,m}\rangle &= |A;s,m\rangle_{NS} - |A;s,m\rangle_R
~}}
and they have vanishing self-overlaps as expected from supersymmetry.

\vskip 15pt
\noindent
\mathboxit{\it class ~1 - class ~2}
\smallskip
\noindent
We conclude this section with a brief survey of the
cylinder amplitudes between class 1 and class 2 branes.
The explicit form of these amplitudes is
\eqn\dad{
_{NS}\langle A|e^{-\pi T H^c}|A;s,m\rangle_{NS} = \frac{1}{2}
\bigg(
\chi_c(s,m;it)\bigg[ {0 \atop 0}\bigg]
\frac{\theta [{0 \atop 0} ](it)}{\eta(it)^3} - \chi_c(s,m+\frac{1}{2};it)\bigg[ {1 \atop 0}\bigg]
\frac{\theta [{1 \atop 0} ](it)}{\eta(it)^3}\bigg)
~,}
\eqn\dae{\eqalign{
_{R}\langle A|e^{-\pi T H^c}|A;s,m\rangle_{R} =
-\frac{1}{2}
\bigg(\chi_c(s,m;it)\bigg[ {0 \atop 1}\bigg]
\frac{\theta [{0 \atop 1} ](it)}{\eta(it)^3}
- \chi_c(s,m+\frac{1}{2};it)\bigg[ {1 \atop 1}\bigg]
\frac{\theta [{1 \atop 1} ](it)}{\eta(it)^3} \bigg)
~.}}
Supersymmetric D-brane configurations can be deduced from
the vanishing amplitudes
\eqn\daea{\eqalign{
\langle A |e^{-\pi T H^c}|A;s, 0\rangle &= \frac{1}{2}\Lambda_{1}(s;it)=0~,
\cr
\langle A|e^{-\pi T H^c}|\overline{A;s,1/2}\rangle &=
\frac{1}{2} \Lambda_{-1}(s;it) = 0
~.}}

\subsec{A brief summary of the proposed D-branes}

In the preceding analysis we considered D-branes in the four-dimensional
non-critical type IIB superstring theory \iae\
that have Neumann boundary conditions in the four flat directions,
varying dimensionality in $SL(2)/U(1)$ and different BPS properties.
D-branes in the type IIA or type IIB theory with lower dimensionality in $\IR^{3,1}$
can be obtained easily by T-duality and will not be discussed here explicitly.

More precisely, we found a D3-brane (denoted by
the boundary state $|A\rangle$) and its anti-brane, both of which are
separately BPS. The worldvolume of this brane is supported near the tip of the cigar.
We also obtained D4- and D5-branes which are extended in the radial direction
of the cigar. Both of these branes are labeled by a non-negative continuous
real parameter $s$ and an extra $\IZ_2$ label $m=0,\frac{1}{2}$.
The B-type, class 2 D4-branes are non-BPS since they couple only to NSNS sector states.
On the other hand, the D5-branes denoted by the boundary state $|A,s,m\rangle$ are BPS.
Geometrically, the D5-branes cover the cigar partially or totally
starting from the asymptotic circle at infinity and terminating
at a finite distance $\rho_{\min} \sim s \geq 0$ from the tip.
The analysis of the corresponding annulus amplitudes revealed that the D5-branes
are double-sheeted, $i.e.$ they have two branches in the T-dual
trumpet geometry.

\newsec{General properties of the BPS branes}

The BPS D3- and D5-branes of the previous section are sources for
the appropriate RR fields of the non-critical theory. In this
section we want to elaborate on the nature of the corresponding RR
couplings and the potential presence of dangerous non-dynamical RR
tadpoles. In the process we also discuss the dictionary between
branes in the non-critical superstring theory and branes in the
corresponding NS5-brane configuration of
\refs{\ElitzurFH,\ElitzurHC,\GiveonSR}.

As explained in section 2, from the six-dimensional point of view
the type IIB theory has RR fields coming from the R$-$R$-$ and R$+$R$+$
sectors. The massless RR potentials are
\eqn\eea{
C_0, ~ C_2^{+}, ~ C_4
~}
and appear only in the R$+$R$+$ sector at the bottom of a continuous spectrum.

In the critical superstring, D3-branes couple electrically to the
four-form potential $C_4$ through the standard WZ coupling
\eqn\eeaa{ \int d^4 x ~ C_4 ~.} In the present non-critical case,
this statement is slightly obscured by the non-trivial profile of
the D3-brane, which extends along the radial direction of the
cigar but is mainly supported near the tip. Furthermore, the class
1 boundary conditions on the free fermions of the theory are
Neumann in all directions; in particular, they are Neumann in both
the radial and the angular directions of the cigar. In other
words, one has to impose on the free fermions the same boundary
conditions as in the case of the class 2 D5-branes, which we
formulated with the use of the same Ishibashi states. In that
sense, it is more appropriate to think of the class 1 D3-branes as
small D5-branes localized near the tip of the cigar. Hence, in
order to understand how they couple to RR fields it helps to
understand first the corresponding couplings of the D5-branes.

In flat spacetime, D5-branes couple electrically to a six-form
potential $C_6$. In the present non-critical case, six dimensions
account for the full dimensionality of spacetime and the six-form
is a non-dynamical field - the analog of the $C_{10}$ potential in
ten-dimensional flat spacetime, whose source is the D9-brane in
type IIB. In ten dimensions a configuration of D9-branes with a
non-vanishing $C_{10}$ tadpole is a serious problem. Such tadpoles
are usually cancelled by introducing orientifold planes or the
appropriate number of anti-D9-branes. Is there a similar $C_6$
tadpole from the D5 boundary states $|A;s,m\rangle$ in the
non-critical case? We would like to argue that the answer to this
question is negative. First of all, the boundary states $|A;s,m\rangle$
describe D5-branes with two sheets of opposite orientation. Asymptotically
in the radial direction of the cigar, this
configuration resembles a brane-antibrane pair and
hence should have a vanishing $C_6$ charge. Despite this feature
this system is supersymmetric and does not exhibit any open string
tachyons. Secondly, the absence of any pathological non-dynamical
tadpoles is expected to mesh nicely with the corresponding picture
in the type IIA NS5-brane configuration, which appears in fig.\ 4.
The correspondence with this configuration is another interesting
aspect of the present discussion and we would to take a minute to
summarize some of the relevant details.

In fig.\ 4, the finite D4-branes suspended along the 6-direction
between the NS5-branes
\eqn\eeg{\eqalign{
NS5 ~ &: ~ (x^0,x^1,x^2,x^3,x^4,x^5)~,
\cr
NS5' ~ &: ~ (x^0,x^1,x^2,x^3,x^8,x^9)
}}
correspond to the class 1 D3-branes of the non-critical superstring
setting. Accordingly, the type IIA D6-branes
\eqn\eei{
D6 ~ : ~ (x^0,x^1,x^2,x^3,x^7,x^8,x^9)
}
have similar characteristics with the D5-branes
$|A;s,0\rangle$, $|\overline{A;s,\frac{1}{2}}\rangle$,
which are T-dual to the D4-branes of fig.\ 6
in the trumpet geometry.\foot{We will say more about
this correspondence in section 5 below. For example, the
$|A;s,0\rangle$ and $|\overline{A;s,\frac{1}{2}}\rangle$ branes exhibit
some important differences.}

{\vbox{{\epsfxsize=115mm
\nobreak
\centerline{\epsfbox{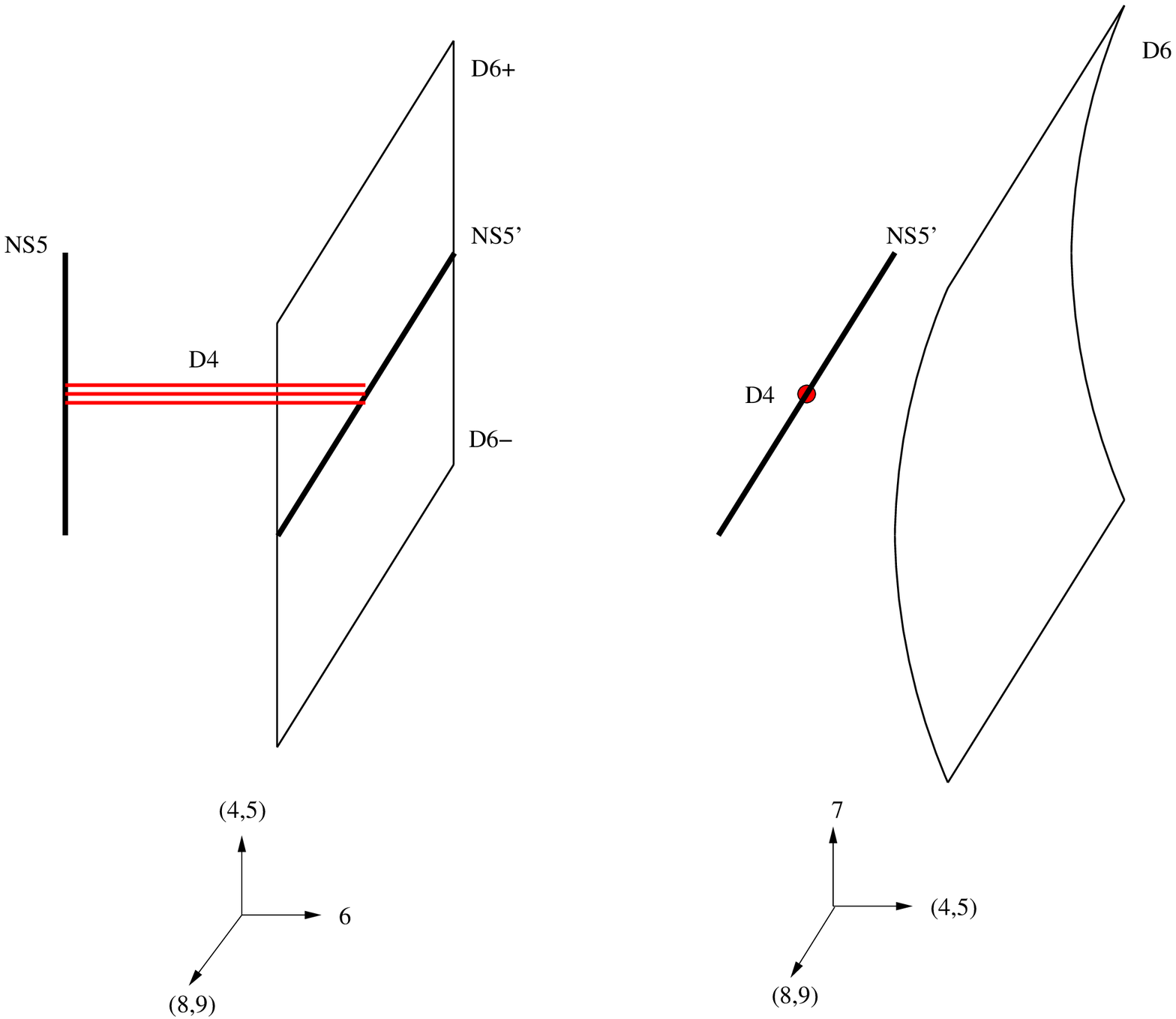}}
\nobreak\bigskip
{\raggedright\it \vbox{
{\bf Figure 4.}
{\it The NS5-brane configuration of fig.\ 1 including D6-branes.
On the left, the NS5'-brane is embedded inside a D6-brane extended in $x^7$.
On the right, the D6-brane has been moved on the $(4,5)$ plane
away from the origin and comes within a minimum distance
from the NS5'-brane without intersecting it.
} }}}}
\medskip}

\medskip
Both the D6-brane of fig.\ 4 and the D4-branes of fig.\ 6 come
from the asymptotic infinity towards the throat and then return
back. When the  D6-branes of fig.\ 4 approach the NS5'-brane they
can intersect it at $x^4=x^5=x^7=0$ (see the figure on the left)
or they can come within a minimum distance of the NS5'-brane at a
locus of points with $x^4, x^5 \neq 0$, and $x^7=0$ (see the
figure on the right). The special situation where the D6-branes
meet the NS5'-brane at $x^7=0$ corresponds to the non-critical
D5-branes with $s=0$. In that case, the upper and lower sheets of
the D6-brane correspond to the two separate sheets of the
D5-brane. Clearly, we do not expect non-dynamical tadpoles in
fig.\ 4 and the same goes for the class 2 D5-branes in the
non-critical superstring theory.

Nevertheless, we still observe that the D5-branes have
a non-zero coupling to massless R$+$R$+$ potentials.
What is the rank of these potentials and how do they couple
to a six-dimensional worldvolume?

On the level of the effective spacetime action there are several
ways that the dynamical RR potentials couple to the D5-branes of
section 3. First of all, it is known \FotopoulosVC\ that D2-branes
on the cigar can have a non-vanishing background gauge field
strength $F_2$ on their worldvolume.\foot{As we are about to see
in the next section, there is no massless gauge field on the
D5-branes, but there is a massless scalar which can be thought of
as the fluctuation of a two-form field strength $F_2$.} This
implies that the spacefilling D5-branes can have WZ couplings of
the form
\eqn\eeb{ \int d^6x ~e^{-\Phi} F_2 \wedge C_4 ~.}
The presence of this coupling indicates that the class 2 branes of
section 3 have an induced D3-brane charge and it would be
interesting to understand its implications for the analysis of
\KlebanovYA.

It is an open question whether there exist any non-trivial WZ
couplings due to the curvature of the cigar. An obvious choice is
a coupling of the form
\eqn\eed{ \int d^6x ~ e^{-\Phi}\tr(R\wedge
R) \wedge C_2^{+} ~.}
We are not aware of an explicit
demonstration of such WZ couplings in the non-critical superstring
case, but it would be useful to derive and verify their presence
with a tree-level calculation on the disc. Analogous statements should apply
also to the D3-brane boundary states, which are based on the same Ishibashi states as the
D5-branes and therefore should couple to the $C_4$ RR potential in a similar fashion.

Finally, a potentially worrying aspect of having a D-brane setup with non-vanishing
D3-brane flux is the following. A D3-brane in our six-dimensional
non-critical setting is similar to a D7-brane in ten-dimensional flat space, which
is pathological. The origin of the pathology lies in the low co-dimension that
does not allow the flux lines to decay appropriately fast in the asymptotic
infinity. For a D7-brane in ten dimensions, the co-dimension is two and
the solution of the Laplace equation in the two-dimensional transverse space is logarithmic
suggesting that we cannot ignore the backreaction of the brane.

At first sight, the same conclusion would seem to hold for a D3-brane
in our six-dimensional space. A more careful examination, however,
shows that this is not the case. The two-dimensional Laplace equation
on the axially-gauged cigar geometry of $SL(2)_1/U(1)$ takes the
form \DijkgraafBA
\eqn\eee{
\bigg[\frac{\d^2}{\d \rho^2}+\coth \rho \frac{\d}{\d \rho}+
\coth^2\frac{\rho}{2}\frac{\d^2}{\d \theta^2}\bigg]T(\rho,\theta)=0
~,}
which becomes
\eqn\eef{
\bigg[\frac{\d^2}{\d \rho^2}+\frac{\d}{\d \rho}
+\frac{\d^2}{\d\theta^2}\bigg]T(\rho,\theta)=0
~}
at the asymptotic region $\rho \rightarrow \infty$.
For wavefunctions of the form $T(\rho,\theta)=f(\rho)e^{im\theta}$
this equation has two solutions for $f(\rho)$, one exponentially growing
and another exponentially decaying. Hence the problem with the logarithmic
divergence does not appear.

\newsec{Four-dimensional gauge theories on D$3$-D$5$ systems}

We are now in position to realize the main purpose of this paper,
which is to obtain four-dimensional $\NN=1$ SQCD as the low-energy
theory of the modes living on a configuration of D-branes in the four-dimensional
non-critical superstring \iae. $\NN=1$ SQCD is an $SU(N_c)$ super-Yang-Mills
theory with $N_f$ flavour chiral superfields $Q^i$ in the fundamental
${\bf N}_c$ of the gauge group and
$N_f$ flavour chiral superfields $\tilde Q_{\tilde i}$ in the anti-fundamental
$\bar {\bf N}_c$ ($i,\tilde i=1,...,N_f$).
For $N_f \leq 3 N_c$ this theory is asymptotically free
and has an infrared behaviour that depends crucially
on $N_c$ and $N_f$. In particular, for
$N_f > N_c + 1$ it exhibits a very interesting
electric-magnetic duality, known as Seiberg-duality \SeibergPQ,
which exchanges the above electric description with a dual magnetic one
that has different ultraviolet properties but the same infrared behaviour.
The classical symmetries and moduli of $\NN=1$ SQCD will be discussed later
in this section, where it will be examined which properties of the gauge theory
can be realized directly in a D-brane setup in non-critical superstring theory.

\subsec{The D-brane setup and the spectrum of open strings}

The SYM part of $\NN=1$ SQCD can be realized
on $N_c$ D3-branes at the tip of the cigar. The spectrum
of $3$-$3$ strings can be deduced from the amplitude
$\langle A|e^{-\pi T H^c}|A\rangle$ in section 3 and
contains massless fields that belong in a $\NN=1$
vector supermultiplet. Indeed, the 3-3 open string spectrum
comprises of a bosonic NS$+$ sector and a fermionic R$-$ sector.
The leading order expansion of the NS$+$ sector character gives
two physical massless modes
\eqn\faa{
\frac{1}{2}\bigg(\chi_I(it)\bigg[{ 0 \atop 0}\bigg] \frac{\theta \big[{ 0 \atop 0}\big](it)}
{\eta(it)^3} - \chi_I(it)\bigg[{ 0 \atop 1}\bigg] \frac{\theta \big[{ 0 \atop 1}\big](it)}
{\eta(it)^3} \bigg) \sim 2 +\OO(q)
~}
and the same result holds for the R$-$ sector as well. This is the
right multiplicity for the physical modes of a four-dimensional gauge field
and the corresponding gauginos. Hence, putting $N_c$ D3-branes
on top of each other gives the full spectrum
of pure $U(N_c)$ super Yang-Mills.\foot{In the D-brane configurations
of Hanany-Witten type the $U(1)$ is frozen in the quantum theory and decouples \WittenSC.
Presumably the same happens also in our case. However,
the quantum properties of the present configurations will not be discussed here,
since they lie outside the immediate scope of this paper.}

One can realize the chiral superfields $Q^i$ and $\tilde Q_{\tilde i}$ with an extra
set of $N_f$ D5-branes. In the language of section 3 these should be A-type
class 2 branes and the available boundary states are
\eqn\fab{
|A;s,m\rangle~, ~ ~ |\overline{A;s,m}\rangle~, ~ ~ s\in \IR_{\geq 0}~,
~ ~ m=0,\frac{1}{2}
~.}
In the presence of D3-branes only the following subset of
boundary states leads to supersymmetric configurations
\eqn\fac{
|A;s,0\rangle~, ~ ~ |\overline{A;s,\frac{1}{2}}\rangle
~.}
Since they are double-sheeted, we expect that $N_f$ branes of this type
will be sufficient in realizing the full matter content of $\NN=1$ SQCD, which
includes an equal number of superfields in the fundamental and
the anti-fundamental.

Indeed, these superfields will arise as the lowest
level excitations of 3-5 strings. In section 3, we presented the annulus amplitudes
\eqn\fad{\eqalign{
\langle A |e^{-\pi T H^c}|A;s, 0\rangle &= \frac{1}{2}\Lambda_{1}(s;it)=0~,
\cr
\langle A|e^{-\pi T H^c}|\overline{A;s,1/2}\rangle &=
\frac{1}{2} \Lambda_{-1}(s;it)= 0
~.}}
Massless excitations of 3-5 strings appear only in the character combination
$\Lambda_{-1}(s;it)$ for the special case $s=0$.
Hence, from now on we concentrate on D5-branes represented
by the boundary state $|\overline{A;s,\frac{1}{2}}\rangle$.
For this choice 3-5 strings include at the lowest level an equal
number of massless NS$-$ bosons and R$+$ fermions,
which form $two$ massless $\NN=1$ chiral multiplets. This can be seen directly
from the character expansion
\eqn\fae{\eqalign{
\Lambda_{-1}(s;it)&=
\Big(\chi_c(s,\frac{1}{2};it)\bigg[ {0 \atop 0}\bigg]
\frac{\theta \big[{0 \atop 0} \big](it)}{\eta(it)^3}+
\chi_c(s,\frac{1}{2};it)\bigg[ {0 \atop 1}\bigg]
\frac{\theta \big[{0 \atop 1} \big](it)}{\eta(it)^3}\Big)
\cr
&-\Big(\chi_c(s,0;it)\bigg[ {1 \atop 0}\bigg]
\frac{\theta \big[{1 \atop 0} \big](it)}{\eta(it)^3}+
\chi_c(s,0;it)\bigg[ {1 \atop 1}\bigg]
\frac{\theta \big[{1 \atop 1} \big](it)}{\eta(it)^3}\Big)
\cr
&=\bigg( 4q^{s^2} + \OO(q^{s^2+\frac{1}{2}})\bigg)_{NS-}-
\bigg(4q^{s^2} + \OO(q^{s^2+\frac{1}{2}})\bigg)_{R+}
~,}}
which is quoted here for arbitrary $s$. Moreover,
using the character identities of
appendix A we can rewrite $\Lambda_{-1}(0;\tau)$
in terms of discrete characters as
\eqn\fag{\eqalign{
\frac{1}{2}\Lambda_{-1}(0;\tau)&=
\bigg(\chi_d(\frac{1}{2},0;\tau)\bigg[{0 \atop 0}\bigg]
\frac{\theta\big[ {0 \atop 0}\big](\tau)}{\eta(\tau)^3}+
\chi_d(\frac{1}{2},0;\tau)\bigg[{0 \atop 0}\bigg]
\frac{\theta\big[ {0 \atop 1}\big](\tau)}{\eta(\tau)^3}\bigg)
\cr
&-\bigg(\chi_d(\frac{1}{2},1;\tau)\bigg[{1 \atop 0}\bigg]
\frac{\theta\big[ {1 \atop 0}\big](\tau)}{\eta(\tau)^3}+
\chi_d(\frac{1}{2},1;\tau)\bigg[{1 \atop 1}\bigg]
\frac{\theta\big[ {1 \atop 1}\big](\tau)}{\eta(\tau)^3}\bigg)
~.}}

It is natural to interpret the two lowest level contributions
in \fag\ as the quark supermultiplets $Q^i$
and $\tilde Q_{\tilde i}^{\dagger}$ respectively. Geometrically, these fields originate from
3-5 strings stretching between the D3-brane and different sheets
of the D5-brane $|\overline{A;0,\frac{1}{2}}\rangle$.
The superfields $Q^i$ appear with momentum $n=\frac{1}{2}$
and transform in the fundamental representation ({\bf N}$_c$, {\bf N}$_f$)
of $U(N_c)\times U(N_f)$. The second set of chiral
superfields $\tilde Q_{\tilde i}$ has the same momentum,
transforms in the anti-fundamental ($\bar{\bf N}_c$, $\bar{\bf N}_f$) and
arises from the opposite orientation 5-3 strings.

The above picture is perfectly consistent with the one expected from
the NS5-brane configuration in fig.\ 4.
In the situation depicted on the left of that figure
the D6-brane splits into two pieces, which we
call D6$+$ and D6$-$. Each of them corresponds to one of the sheets
of the class 2 D5-brane $|\overline{A;0,\frac{1}{2}}\rangle$. Strings stretching between
the D4-branes and D6$+$ are expected to give rise to the quark supermultiplets
$Q^i$, whereas strings stretching between the D4-branes and D6$-$
are expected to give rise to the quark supermultiplets $\tilde Q_{\tilde i}^{\dagger}$
\refs{\HananySA,\ElitzurPQ}.

Consequently, in what follows we consider a setup
of $N_c$ D3-branes and $N_f$ D5-branes
described respectively by the boundary states $|A\rangle$ and
$|\overline{A;0,\frac{1}{2}}\rangle$
and we argue that they realize the electric description of $\NN=1$ SQCD.
The r\^ole of D5-branes with $s>0$, will be clarified shortly.
Note that the other class of D5-branes represented by the boundary state
$|A;s,0\rangle$ gives massive 3-5 spectra in the NS$+$, R$-$ sectors and
does not appear to play a r\^ole, when we try to engineer $\NN=1$ SQCD.

So far we have discussed the spectrum of 3-3 and 3-5 strings. Now
we turn to the spectrum of 5-5 strings. This can be read off the
annulus amplitude
\eqn\fai{\eqalign{ \bigg\langle
\overline{A;0,\frac{1}{2}}\bigg|e^{-\pi T H^c}
\bigg|\overline{A;0,\frac{1}{2}}\bigg\rangle &= \int_0^{\infty}
ds' ~ \big[ \big(\rho_1(s';0|0)+\rho_2(s';0|0)\big)
\Lambda_1(s';it) \cr &+\big(\rho_1(s';0|0)-\rho_2(s';0|0)\big)
\Lambda_{-1}(s';it)\big] ~,}}
where $\rho_1$, $\rho_2$ are the
spectral densities of eqs.\ \densityinteger, \densityhalfinteger.
The most notable characteristics of this spectrum are the
following. First, it does not exhibit any massless vector
multiplets, which would correspond to massless gauge fields on the
D5-branes. Vector multiplets appear in the NS$+$ and R$-$ sectors,
which are captured by the $\Lambda_1(s;\tau)$ character. There are
no massless contributions to this character for any value of $s$.
Although the existence of a massive vector seems strange
at first sight, it is a natural characteristic of linear
dilaton backgrounds.

The second notable characteristic of
the spectrum \fai\ is a massless chiral multiplet $M_i^{\tilde j}$ in the
bifundamental of $U(N_f)\times U(N_f)$ with quantum numbers $s=0$, $m=1/2$.
This mode has a natural superpotential coupling to the quarks $Q^i$, $\tilde Q_{\tilde j}$
\eqn\faj{
W_M=\Tr M_i^{\tilde j} Q^i \tilde Q_{\tilde j}
~,}
which can be deduced from the respective three-string tree-level interaction.
Notice that a similar coupling appears in the magnetic description of SQCD
for the elementary magnetic mesons. Hence, one may wonder whether
we are really discussing the magnetic description of SQCD and
if we should interpret the massless multiplets $M_i^{\tilde j}$
as the magnetic mesons of that description. However, the fact that
the multiplets $M_i^{\tilde j}$ appear at the bottom
of a continuous spectrum with arbitrary radial momentum in the cigar
direction indicates that they do not constitute propagating UV degrees of freedom
in the D3-brane gauge theory. Instead, they should
be regarded as parameters in this gauge theory. The precise meaning
of these parameters in the electric description of SQCD is the following.

\bigskip
{\vbox{{\epsfxsize=35mm
\nobreak
\centerline{\epsfbox{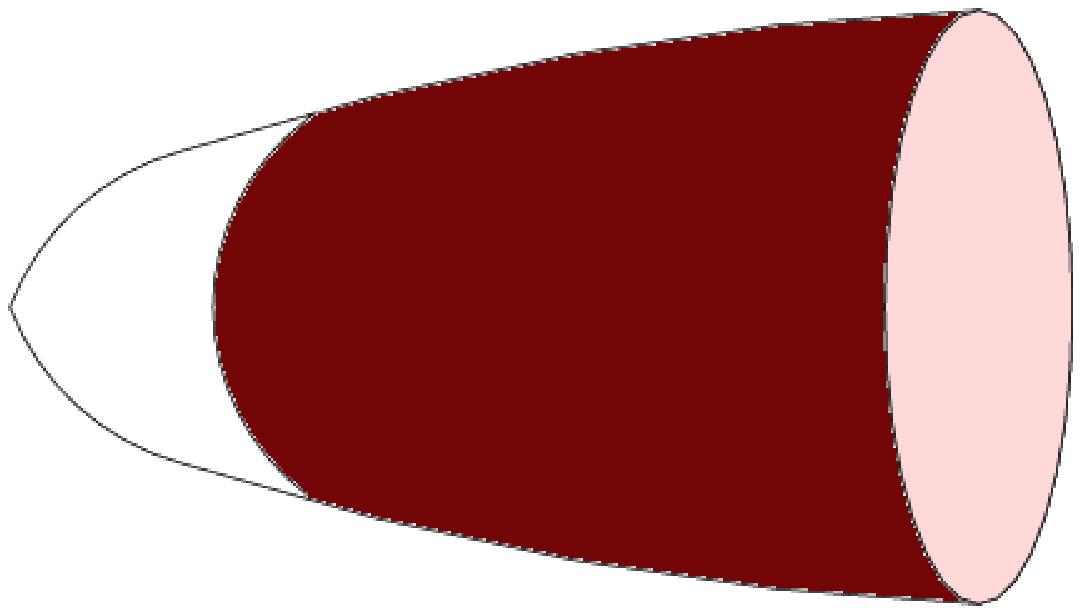}}
\nobreak\bigskip
{\raggedright\it \vbox{
{\bf Figure 5.}
{\it The geometric picture of a cigar D2-brane corresponding to
a class $2$ boundary state. It covers the cigar partially
up to a minimum distance $s$ from the tip.
} }}}}
\bigskip}

The superpotential coupling \faj\ implies that
vacuum expectation values (vev's) of the $M_i^{\tilde j}$ operators
give masses to the quarks $Q$, $\tilde Q$ and generate (a subset of)
the usual mass deformations of $\NN=1$ SQCD. These deformations have
a clear geometric meaning in our setup that can be
understood by considering more closely the worldvolume
theory of the flavor branes. We can see directly from
equations \fad\ and  \fae\ that turning on the mass parameter
$M_i^{\tilde i}$ for the single $i$th D5-brane corresponds to shifting the modulus $s$ of
the class 2 branes by an amount $s_i$ proportional to $|M_i^{\tilde i}|$.\foot{The existence
of this mass deformation is another reason to expect two chiral multiplets
in the spectrum of 3-5 strings and substantiates the validity of the normalization
of the class 1 and class 2 boundary states in section 3.
A single chiral multiplet cannot give rise to a holomorphic gauge-invariant
mass deformation.} Hence, by turning on this deformation we expect to
get the class 2 boundary state of a D5-brane that wraps the cigar and extends
from the asymptotic infinity up to a distance $s_i$ from the tip (see fig.\ 5).
Notice that in this process the two sheets of a single flavour brane
cannot move independently and
we can only obtain the diagonal vev's
\eqn\fak{
M_i^{\tilde j}=m_i \delta_i^{\tilde j}
~}
(no summation implied).
Each vev $m_i$ is in one-to-one
correspondence with the single modulus $s_i$ of the class 2 D5-branes
\eqn\faka{
\big|\overline{A;s_i,\frac{1}{2}}\big\rangle
~.}

\subsec{Symmetries and moduli}

At this point, we want to make a few general comments about the
classical symmetries and moduli of $\NN=1$ SQCD\foot{The quantum moduli space is
not accessible to our tree-level classical (type II) string theory description.}
and see if and how they can be realized geometrically in the D-brane configurations
of this paper. In the absence of a superpotential, the classical symmetry of the theory
is
\eqn\fama{
SU(N_f)_L\times SU(N_f)_R \times U(1)_B \times U(1)_a \times U(1)_x
~.}
The two $SU(N_f)$ factors rotate the chiral multiplets $Q^i$, $\tilde Q_{\tilde j}$.
$U(1)_B$ is a vector-like baryon symmetry, which assigns charge $+1$ $(-1)$
to $Q$ ($\tilde Q$).
$U(1)_a$ and $U(1)_x$ are R-symmetries under which the gaugino has
charge one and the quarks $Q$, $\tilde Q$ have charge 1 or 0. Quantum mechanically
only one combination of the two R-symmetries is anomaly free.

The vector $SU(N_f)$ global symmetry is present in any
configuration with the same parameters $s_i$ for all flavor branes.
The appearance of a second axial $SU(N_f)$, when all the matter
multiplets are massless, can be seen more easily in the T-dual trumpet geometry.
The mass deformed theory involves D5-branes with parameters $s>0$,
which look like the D1-branes of fig.\ 6 in the T-dual trumpet.
The two D1-branches are connected to each other and therefore
exhibit a single (vector) $SU(N_f)$ symmetry.
For $s=0$ however, the two branches
are disconnected and go straight into the
strong coupling singularity. Then we can associate
an $SU(N_f)$ symmetry to each one of the two independent branches, leading
to an enhancement of the flavour symmetry to $SU(N_f) \times SU(N_f)$.
This reproduces exactly  the field theory result \fama.
Similar statements in the context of the NS5-brane configuration in fig.\ 4
can be found in \refs{\BrodieSZ,\HananySA,\GiveonSR}.

The closed string theory on the cigar has three conserved
$U(1)$ currents \MurthyES. Two of them
are the chiral and anti-chiral $\NN=2$ currents
$J_{{\cal N}=2}$ and  $\bar J_{{\cal N}=2}$, and the third is
the non-chiral current associated to
the momentum in the angular direction of the cigar.
Only the last current commutes with the BRST symmetry
and constitutes a physical current of the non-critical
superstring theory.\foot{We would like to thank S.\ Ashok, S.\ Murthy
and J.\ Troost for pointing this out and for correcting an erroneous
statement in the first version of this paper.}

\bigskip
{\vbox{{\epsfxsize=35mm
\nobreak
\centerline{\epsfbox{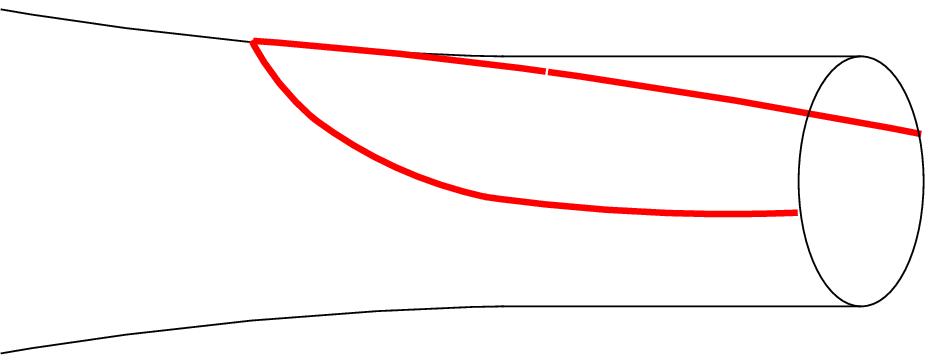}}
\nobreak\bigskip
{\raggedright\it \vbox{
{\bf Figure 6.}
{\it A D1-brane on the T-dual trumpet. The D1-brane comes from the
asymptotic infinity, curves and then turns back to the asymptotic infinity
at the diametrically opposite point.
} }}}}
\bigskip}

Open string theory on the previously discussed D3-D5 configuration
exhibits the following three $U(1)$ symmetries. The first, which
will be called $U(1)_m$ (with charge $Q_m$) is the open string
version of the $U(1)$ angular momentum symmetry of the cigar. The
other two, which will be called $U(1)_L$ and $U(1)_R$ are the
$U(1)$ global symmetries, which are part of the $U(N_f)_L\times
U(N_f)_R$ global flavor symmetry group. We will denote the
corresponding charges as $Q_L$ and $Q_R$. It is useful to define
the following linear combinations of these charges
\eqn\fana{\eqalign{
Q_{\pm}&=\frac{1}{2}(Q_L\pm Q_R)~, \cr
Q_x&=\frac{4}{3}(Q_m+Q_-)~, \cr
Q_a&=\frac{4}{3}(Q_m-\frac{1}{2}Q_-)~, \cr
Q_B&=-2Q_+~. }}
The
three charges $Q_B, Q_a$ and $Q_x$ are in one-to-one
correspondence with the SQCD $U(1)$ symmetries $U(1)_B, U(1)_a$
and $U(1)_x$. Indeed, $Q_B$ is the charge of a global $U(1)$
symmetry and $Q_x, Q_a$, both of which involve the $U(1)_m$
charge, are the charges of two $U(1)_R$ symmetries.

One can check that with this identification the $U(1)$ charge assignments
work as expected from SQCD. The chiral superfield $M$ has
$U(1)_B\times U(1)_a \times U(1)_x$ charges $(0,0,2)$ and
the quark superfields $Q$ and $\tilde Q$ have
respectively $(1,1,0)$ and $(-1,1,0)$.

The geometric interpretation of the $U(1)_R$ symmetries $U(1)_a$
and $U(1)_x$ is most clear in the T-dual trumpet background of
fig.\ 6. In that case, the momentum symmetry $U(1)_m$ becomes a
winding symmetry $U(1)_w$. The charges $Q_a$ and $Q_x$ measure
respectively the winding of open strings on the lower and upper
half of the asymptotic cylinder in fig. 6.\foot{Strictly speaking, this
is true for the linear combinations

\centerline{
$Q'_a=Q_m-\frac{1}{2}Q_-~, ~ ~ Q'_x=2(Q_m+\frac{1}{2}Q_-)$~.}
}
For D4-branes with $s>0$ $Q_a$ is conserved, but $Q_x$ is not. Pictorially,
strings winding around the bottom half of the asymptotic cylinder
cannot unwind by reaching the turning point of the brane, but
strings winding around the upper half can.\foot{We would like
to thank S.~Murthy, who suggested this picture to us.} This fits nicely with
the fact that the superfield $M$ is charged under $U(1)_x$, but
uncharged under $U(1)_a$. As a result, non-zero vev's of $M$ break
$U(1)_x$ explicitly, but they preserve $U(1)_a$.

We would like to finish with a few comments  on the classical moduli space
of $\NN=1$ SQCD. It is well-known that the dimensionality
of this space depends crucially on
the number of colours and flavours $N_c$ and $N_f$ respectively.
For $N_f<N_c$ the moduli space is $N_f^2$ dimensional and can
be labeled by the gauge invariant meson fields $Q^i \tilde Q_{\tilde j}$.
By giving non-zero vev's to the massless quarks $Q$, $\tilde Q$ one
can Higgs the gauge group down to $SU(N_c-N_f)$.
For $N_f\geq N_c$ new gauge invariant baryon fields appear
and the dimension of the moduli space becomes $2N_cN_f-N_c^2$.
The gauge group can now be broken completely by the Higgs mechanism.

In the brane description of \ElitzurHC, Higgsing corresponds to splitting
fourbranes on sixbranes in the presence of NS5-branes according to the
$s$-rule. In our setting, Higgsing
corresponds to a marginal deformation of the open string theory living
on our D3-D5 setup, but this deformation does not appear to have
an obvious geometric meaning. If Higgsing can be described
geometrically in our setup, it would imply a non-trivial statement
involving the D3-branes at the tip of the cigar. Obviously, it would be
extremely interesting to understand this point better.
Among other things, this could be useful for
a microscopic derivation of the ``phenomenological'' $s$-rule
of D-brane dynamics in the vicinity of NS5-branes. We hope to
return to this interesting issue in future work.

\newsec{Future prospects}

In this paper we studied several aspects of D-brane dynamics in
a specific four-dimensional non-critical superstring theory,
which involves the $\NN=2$ Kazama-Suzuki model for $SL(2)/U(1)$ at level $1$.
D-branes in this theory were treated with exact boundary conformal field
theory methods building on previous work on the
$\NN=2$ Liouville theory and $\NN=2$ Kazama-Suzuki model with boundary
\refs{\RibaultSS\EguchiIK\AhnTT\IsraelJT\AhnQB\FotopoulosUT-\HosomichiPH}.
A similar analysis for the more general case \iaa\ can be performed
with analogous techniques and it will be useful for a better understanding
of D-brane dynamics in closely related situations
involving non-critical superstring theory, string theory in the vicinity
of Calabi-Yau singularities, and the near-horizon geometry of NS5-branes.
In general, this study is expected to yield interesting information about
gauge theories and LSTs. Related work in this direction has appeared recently in
\IsraelFN.

Our primary goal in this paper was to understand some of the key features
of the general story by studying a specific example
that realizes $\NN=1$ SQCD. There are several aspects of
our analysis that deserve further study.
For example, it would be very interesting to see if we can
obtain the dual magnetic description of SQCD using D-branes
in the non-critical superstring \iae. This seems difficult
to achieve solely with the D-branes presented in section 2.
On the other hand, the general analysis of D-branes in
the background of NS5-branes \`a la Hanany-Witten suggests that this
should be possible. If so, can we also understand
Seiberg duality as a classical statement of the corresponding
D-brane configurations? Within the framework of NS5-brane
setups \refs{\ElitzurHC,\GiveonSR}, or within its
T-dual involving Calabi-Yau singularities \OoguriIH,
there are convincing arguments that demonstrate Seiberg duality in this
way.

Another interesting question is whether the Higgs moduli of
$\NN=1$ SQCD have a clear geometric meaning in terms of D-brane
configurations in the non-critical superstring description. This would
be a non-trivial statement involving the D3-branes at the tip of the cigar
and may also lead to a microscopic derivation or at least further insight
on the ``phenomenological'' $s$-rule of D-brane dynamics in
the background of NS5-branes.

Finally, it would be extremely interesting to see whether we can
obtain a better grasp of a generalized AdS/CFT correspondence
within non-critical superstring theory along the lines of \KlebanovYA.
This would open up the road for
a direct analysis of the strong coupling dynamics of the class of gauge theories
that can be realized in non-critical superstring theory and the corresponding
NS5-brane configurations. Clearly, one of the major tasks is to
determine the backreaction of the D3- and D5-branes
on the cigar geometry. A first step in this direction, using supergravity methods,
has been taken in previous work \KlebanovYA\ by Klebanov and Maldacena.
They found a highly curved supergravity solution,
which is relevant for $\NN=1$ SQCD at the conformal window.
A better understanding of this solution, $e.g.$ in
relation to its stability and Seiberg duality, can perhaps be obtained
using the results presented here. For example, calculating the one-point function
of massless closed string fields on the disc and their profile in
the asymptotic infinity is a first exercise that can be done in a straightforward way
using the results of this paper \progress. Of course, in order to proceed further
one would have to compute and resum an infinite set of contributions coming
from higher open string loops (see \BertoliniJY\ for a similar analysis in the critical case).
Also, going beyond supergravity is bound to bring in
the complications due to RR fields.
It would be interesting to see how far one can go
and how useful it is to think about AdS/CFT within the setting of
non-critical superstring theory.

\bigskip

\centerline{\bf Acknowledgements}

We would like to thank I.\ Antoniadis, I.\ Bakas, J.\ P.\
Derendinger, P.\ Di Vecchia, T.\ Eguchi, M.\ Gaberdiel, E.\
Kiritsis, H.\ Klemm, D.\ L\"ust, N.\ Obers, A.\ Paredes, M.\
Petropoulos, C. \ Scrucca, M. \ Serone, Y.\ Sugawara, and A.\
Zaffaroni for useful discussions and correspondence. We are also
grateful to D.\ Kutasov for various comments on the manuscript and
useful correspondence and to S.\ Ashok, S.\ Murthy and J.\ Troost
for useful comments on the first version of this paper. The work
of A.F.\ has been supported by a ``Pythagoras'' Fellowship of the
Greek Ministry of Education and partially supported by INTAS
grant, 03-51-6346, CNRS PICS \# 2530, RTN contracts,
MRTN-CT-2004-512194, MRTN-CT-2004-0051104 and MRTN-CT-2004-503369,
and by a European Union Excellence Grant MEXT-CT-2003-509661. The
work of N.P. has been supported by the Swiss National Science
Foundation and by the Commission of the European Communities under
contract MRTN-CT-2004-005104.

\appendix{A}{Useful Formulae}

\subsec{Useful identities}

For quick reference, we quote here a few identities involving
the characters of discrete representations. First of all, one
can show that the continuous characters for $s=0$ can be
written as
\eqn\appba{
\chi_c(s=0,\frac{a+1}{2};\tau,z)\bigg[ {a \atop b}\bigg]=
\chi_d(\frac{1}{2},\frac{a}{2};\tau,z)\bigg[ {a \atop b}\bigg]+ (-)^b
\chi_d(1,\frac{a}{2};\tau,z)\bigg[ {a \atop b}\bigg]
~.}
With the use of the identity
\eqn\appbb{
\chi_d(1,\frac{a}{2};\tau,-z)\bigg[ {a \atop b} \bigg]=
(-)^{b+ab}\chi_d(\frac{1}{2},\frac{a}{2};\tau,z)\bigg[ {a \atop b} \bigg]
}
we can also write eq.\ \appba\ as
\eqn\appbc{
\chi_c(s=0,\frac{a+1}{2};\tau,z)\bigg[ {a \atop b}\bigg]=
(1+(-)^{ab}) ~ \chi_d(\frac{1}{2},\frac{a}{2};\tau,z)\bigg[ {a \atop b} \bigg]
~.}

In the main text we also define the vanishing character combinations
\eqn\appbd{\eqalign{
\Lambda_{1}(s;\tau)&=
\Big(\chi_c(s,0;\tau,0)\bigg[ {0 \atop 0}\bigg]
\frac{\theta \big[{0 \atop 0} \big](\tau,0)}{\eta(\tau)^3}-
\chi_c(s,0;\tau,0)\bigg[ {0 \atop 1}\bigg]
\frac{\theta \big[{0 \atop 1} \big](\tau,0)}{\eta(\tau)^3}\Big)
\cr
&-\Big(\chi_c(s,\frac{1}{2};\tau,0)\bigg[ {1 \atop 0}\bigg]
\frac{\theta \big[{1 \atop 0} \big](\tau,0)}{\eta(\tau)^3}-
\chi_c(s,\frac{1}{2};\tau,0)\bigg[ {1 \atop 1}\bigg]
\frac{\theta \big[{1 \atop 1} \big](\tau,0)}{\eta(\tau)^3}\Big)
\equiv 0 ~,}}
\eqn\appbe{\eqalign{
\Lambda_{-1}(s;\tau)&= \Big(\chi_c(s,\frac{1}{2};\tau,0)\bigg[ {0 \atop 0}\bigg]
\frac{\theta \big[{0 \atop 0} \big](\tau,0)}{\eta(\tau)^3}+
\chi_c(s,\frac{1}{2};\tau,0)\bigg[ {0 \atop 1}\bigg]
\frac{\theta \big[{0 \atop 1} \big](\tau,0)}{\eta(\tau)^3}\Big)
\cr
&-\Big(\chi_c(s,0;\tau,0)\bigg[ {1 \atop 0}\bigg]
\frac{\theta \big[{1 \atop 0} \big](\tau,0)}{\eta(\tau)^3}+
\chi_c(s,0;\tau,0)\bigg[ {1 \atop 1}\bigg]
\frac{\theta \big[{1 \atop 1} \big](\tau,0)}{\eta(\tau)^3}\Big)
~\equiv 0 .}}
Using \appba\ and then \appbb\ we can recast $\Lambda_{-1}(0;\tau)$ into the form
\eqn\appbf{\eqalign{
\Lambda_{-1}(0;\tau)&=
\bigg \{ \bigg(\chi_d(\frac{1}{2},0;\tau)\bigg[{0 \atop 0}\bigg]
\frac{\theta\big[ {0 \atop 0}\big](\tau)}{\eta(\tau)^3}+
\chi_d(\frac{1}{2},0;\tau)\bigg[{0 \atop 0}\bigg]
\frac{\theta\big[ {0 \atop 1}\big](\tau)}{\eta(\tau)^3}\bigg)-
\cr
&-\bigg(\chi_d(\frac{1}{2},1;\tau)\bigg[{1 \atop 0}\bigg]
\frac{\theta\big[ {1 \atop 0}\big](\tau)}{\eta(\tau)^3}+
\chi_d(\frac{1}{2},1;\tau)\bigg[{1 \atop 1}\bigg]
\frac{\theta\big[ {1 \atop 1}\big](\tau)}{\eta(\tau)^3}\bigg)\bigg\}+
\cr
&+\bigg\{ \bigg(\chi_d(1,0;\tau)\bigg[{0 \atop 0}\bigg]
\frac{\theta\big[ {0 \atop 0}\big](\tau)}{\eta(\tau)^3}-
\chi_d(1,0;\tau)\bigg[{0 \atop 0}\bigg]
\frac{\theta\big[ {0 \atop 1}\big](\tau)}{\eta(\tau)^3}\bigg)-
\cr
&-\bigg(\chi_d(1,1;\tau)\bigg[{1 \atop 0}\bigg]
\frac{\theta\big[ {1 \atop 0}\big](\tau)}{\eta(\tau)^3}-
\chi_d(1,1;\tau)\bigg[{1 \atop 1}\bigg]
\frac{\theta\big[ {1 \atop 1}\big](\tau)}{\eta(\tau)^3}\bigg)\bigg\}
\cr
&=2\bigg\{ \bigg(\chi_d(1,0;\tau)\bigg[{0 \atop 0}\bigg]
\frac{\theta\big[ {0 \atop 0}\big](\tau)}{\eta(\tau)^3}-
\chi_d(1,0;\tau)\bigg[{0 \atop 0}\bigg]
\frac{\theta\big[ {0 \atop 1}\big](\tau)}{\eta(\tau)^3}\bigg)-
\cr
&-\bigg(\chi_d(1,1;\tau)\bigg[{1 \atop 0}\bigg]
\frac{\theta\big[ {1 \atop 0}\big](\tau)}{\eta(\tau)^3}-
\chi_d(1,1;\tau)\bigg[{1 \atop 1}\bigg]
\frac{\theta\big[ {1 \atop 1}\big](\tau)}{\eta(\tau)^3}\bigg)\bigg\}
~.}}

\subsec{$\SS$-modular transformation properties of the extended characters}

Under the modular transformation $\SS: ~ \tau \rightarrow -\frac{1}{\tau}$
the extended characters presented in the main text transform in the following way
(see for example \EguchiIKK):
\eqn\appaa{\eqalign{
\chi_{c}(s,m;-\frac{1}{\tau},\frac{z}{\tau})
\bigg[ {a \atop b} \bigg]=& 2 (-i)^{ab}
e^{3\pi i z^2/\tau} \sum_{m' \in \IZ_2} e^{-2\pi i m m'}
\int_0^{\infty} ds' \cos(4\pi s s')
\cr
&\chi_{c}(s',\frac{m'}{2};\tau,z)\bigg[ {b \atop a} \bigg]
~,}}
\eqn\appab{\eqalign{
\chi_{d}(j,\frac{a}{2};-\frac{1}{\tau},\frac{z}{\tau})\bigg[ {a \atop b} \bigg]&=
(-i)^{ab}e^{3\pi i z^2/\tau}(-1)^{2bj}
\cr
&~ \bigg[ \int_0^{\infty} ds (-)^b\bigg\{ \chi_{c}(s,0;\tau,z) \bigg[ {b \atop a} \bigg]
-(-)^a\chi_{c}(s,\frac{1}{2};\tau,z)\bigg[ {b \atop a} \bigg]\bigg\}
\cr
&+\frac{i}{2}(-)^{2j}(-)^{ab}\bigg\{ (-)^a
\chi_{d}(\frac{1}{2},\frac{b}{2};\tau,z)\bigg[ {b \atop a} \bigg]
-\chi_{d}(1,\frac{b}{2};\tau,z)\bigg[ {b \atop a} \bigg]\bigg\}\bigg]
~.}}
Using \appbb\ this modular identity can be recast into a simpler form
\eqn\appaba{\eqalign{
\chi_{d}(j,\frac{a}{2};-\frac{1}{\tau},\frac{z}{\tau})\bigg[ {a \atop b} \bigg]&=
(-i)^{ab}e^{3\pi i z^2/\tau}(-1)^{2bj}
\cr
&~ \int_0^{\infty} ds (-)^b\bigg\{ \chi_{c}(s,0;\tau,z) \bigg[ {b \atop a} \bigg]
-(-)^a\chi_{c}(s,\frac{1}{2};\tau,z)\bigg[ {b \atop a} \bigg]\bigg\}
\cr
&- i ~\delta_{ab,1} ~ (-)^{2j} (-)^a \chi_d(\frac{1}{2},\frac{b}{2};\tau,z)
\bigg[ {b \atop a}\bigg]
~.}}
Finally, for the identity characters we have
\eqn\appac{\eqalign{
\chi_{I}(-\frac{1}{\tau},\frac{z}{\tau})\bigg[ {a \atop 0} \bigg]=& 2 (-i)^{ab}
e^{3\pi i z^2/\tau} \int_0^{\infty} ds ~ \sinh(2\pi s)
\cr
& \bigg\{ \tanh(\pi s) \chi_{c}(s,0;\tau,z)\bigg[ {0 \atop a} \bigg]
+(-)^{a} \coth(\pi s) \chi_{c}(s,\frac{1}{2};\tau,z)\bigg[ {0 \atop a} \bigg]
\bigg\}
~,}}
\eqn\appad{\eqalign{
\chi_{I}(-\frac{1}{\tau},\frac{z}{\tau})\bigg[ {a \atop 1} \bigg]=& 2 (-i)^{ab}
e^{3\pi i z^2/\tau} \int_0^{\infty} ds ~ \sinh(2\pi s)
\cr
& \bigg\{ \coth(\pi s) \chi_{c}(s,0;\tau,z)\bigg[ {1 \atop a} \bigg]
+(-)^{a} \tanh(\pi s) \chi_{c}(s,\frac{1}{2};\tau,z)\bigg[ {1 \atop a} \bigg]
\bigg\}
~.}}

\

\subsec{$\SS$-modular transformation properties of classical $\theta$-functions}

The standard definition of theta-functions is
\eqn\appaeaa{
\theta\bigg[ {a \atop b} \bigg](\tau,z)= (-i)^{ab}
\sum_{n=-\in \infty}^{\infty} (-)^{bn} q^{(n-a/2)^2/2} z^{n-a/2}
~.}
Under the transformation $\SS:\tau \rightarrow -\frac{1}{\tau}$
these characters transform as
\eqn\appaea{
\theta\bigg[ {a \atop b} \bigg](-\frac{1}{\tau},\frac{z}{\tau})=(-i)^{ab}
(-i\tau )^{1/2} e^{\pi i z^2/\tau} \theta\bigg[ {b \atop a} \bigg](\tau,z)
~.}
The Dedekind eta function is
\eqn\appaeb{
\eta(\tau)=q^{1/24}\prod_{m=1}^{\infty}(1-q^m)
}
and transforms in the following way
\eqn\appaec{
\eta(-\frac{1}{\tau})=(-i\tau)^{1/2}\eta(\tau)
~.}

\appendix{B}{Chiral GSO projection and the type II torus partition sum}

In this Appendix we review  the
chiral GSO projection that leads to the non-critical superstring partition sum \appggd.
We start by writing down the four-dimensional spin fields
\eqn\sff{
S_{s_0, s_1} = e^{\frac{i}{2}(s_0 H_0+s_1 H_1)}
~,}
where $H_0, H_1$ are the bosonized spacetime fermions and
$s_0, s_1 = \pm \frac{1}{2}$. It is also useful to bosonize
the total ${\cal N}=2$ current with a canonically normalized boson $Y$ so that
\eqn\sffa{
J_{{\cal N}=2}=i \sqrt{\hat c} \partial Y = i \sqrt{3}\partial Y
~.}
We focus only on the case of interest $\hat c=3 \Leftrightarrow k=1$.

The type II non-critical superstring has two sets of spacetime supercharges
\refs{\KutasovUA, \KutasovPV}. One set originates from left-moving fields
on the worldsheet and the other from right-moving fields.
The spacetime supercharges coming from left-moving fields read
\eqn\ssa{
Q^+_{\frac{1}{2},\frac{1}{2}}=
\oint dz \;e^{\frac{1}{2}(-\varphi+i \sqrt{3} Y)} S_{\frac{1}{2} ,\frac{1}{2}},
\;\;\;\;\;\;\;\;
Q^+_{-\frac{1}{2}, -\frac{1}{2}}=
\oint dz\;e^{\frac{1}{2}(-\varphi+i \sqrt{3} Y)} S_{-\frac{1}{2} ,-\frac{1}{2}}
}
\eqn\ssb{
Q^-_{\frac{1}{2}, -\frac{1}{2}}=
\oint dz \;e^{\frac{1}{2}(-\varphi-i \sqrt{3} Y)} S_{\frac{1}{2}, -\frac{1}{2}},
\;\;\;\;\;\;\;\;
Q^-_{\frac{1}{2}, -\frac{1}{2}}=
\oint dz \;e^{\frac{1}{2}(-\varphi-i \sqrt{3} Y)} S_{-\frac{1}{2} ,\frac{1}{2}}
}
where $\varphi$ bosonizes the superghost $\beta, \gamma$ system.
These supercharges are components of a six-dimensional spinor
in the $\bf 4$ of $SO(5,1)$, which can be decomposed as follows
\eqn\ssc{
{\bf 4} \rightarrow {\bf 2}_1 \oplus {\bf \bar 2}_{-1}
}
under the decomposition $SO(5,1) \rightarrow SO(3,1) \times SO(2)$.
Hence, in four dimensions we obtain a Majorana spinor in the $\bf 2 \oplus \bf \bar 2$
of $SO(3,1)$ yielding $N=1$ spacetime supersymmetry.
A similar set of spinors will arise from right-moving fields.
More precisely, for the right-movers we have the option of choosing either the $\bf 4$
or the $\bf 4'$ corresponding to type IIB or type IIA non-critical superstring theory respectively.
In four dimensions, both choices result in a four-dimensional Majorana spinor
$\bf 2 \oplus \bf \bar 2$, since
\eqn\ssd{
{\bf 4}' \rightarrow {\bf 2}_{-1} \oplus {\bf \bar 2}_1
~.}
The overall counting of supercharges yields a theory with $\NN=2$ supersymmetry
in four dimensions. This meshes nicely with the fact that this non-critical string theory
describes holographically a four-dimensional LST on a configuration of tilted
NS5-branes or string theory near a conifold singularity, both of which
preserve 1/4 of the ten-dimensional type II supersymmetry.

On the level of vertex operators the GSO projection requires
locality of all vertex operators with respect to the supercharges.
For a vertex operator of the form \eqn\sse{ \exp((-1+a/2) \varphi
+  i s_0 H_0 + i s_1 H_1 + i Q_a (Y/\sqrt{3})) } this requirement
yields the following integrality condition \eqn\gso{ J_{\rm
GSO}=-1+\frac{a}{2}+(s_0+s_1)+Q_a \;\in\ 2 \IZ ~.} $a=0$ in the
NS-sector and $1$ in the R-sector. For ${\cal N}=2$
primaries\foot{For simplicity, we concentrate here only on the
continuous representations. The discrete representations can be
treated in the same way.} the total $U(1)_R$ charge reads
\eqn\ssf{ Q_a=2\Big(m+\frac{a}{2}\Big)+\frac{a}{2} ~,} where $m$
is the $J^3$ charge of the corresponding bosonic $SL(2)/U(1)$
representation. The two $a$-dependent shifts in $Q_a$ appear,
because $J_{{\cal N}=2}=\psi^+ \psi^- + \frac{2}{k}{\cal J}^3$ and
${\cal J}^3= J^3+\psi^+\psi^-$ is the global $U(1)$ charge that we
gauge in the supersymmetric $SL(2)/U(1)$. Sometimes, it is
convenient to denote the eigenvalue of ${\cal J}^3$ by a separate
parameter $m_t=m+a/2$. Then, we can write $J_{\rm GSO} = F +
2m_t$, where $F=-1 + a/2 +s_0+s_1+a/2$ is the total fermion number
(including the superghost contribution).

In order to obtain a GSO invariant partition function we insert
the projectors
\eqn\ssg{
\frac{1}{2}\big(1+(-1)^{J_{\rm GSO}}\big)~, ~ ~
\frac{1}{2}\big(1+(-1)^{\bar J_{\rm GSO}}\big)
}
inside the trace over the full Hilbert space $\HH$ of the theory. This includes
the 3+1-dimensional flat part, the supersymmetric coset and the ghosts.
Hence,
\eqn\totalpf{
Z_{\rm II} = {\rm Tr}_{\HH} \Bigg( \frac{1+(-1)^{J_{\rm GSO}}}{2}
\;\frac{1+(-1)^{\bar J_{\rm GSO}}}{2}\;
q^{L_0} \bar q^{\bar L_0} \Bigg)
~.}
As usual, the contribution of two of the bosonic (fermionic) degrees of freedom
is cancelled by the contribution of the ghosts (superghosts) and the trace
ends up summing over the two transverse flat directions and the coset.

Let us consider this trace more closely. First, it is instructive
to consider the trace without any GSO projector insertions. Taking
into account the conditions on the NS-sector coset momenta, coming
from the path integral construction of the coset partition
function, $i.e.$ the conditions $m-\bar m = 0$ and $m +\bar m =  w
\; \in \IZ_2$, and obtaining the R-sector by 1/2-spectral flow,
gives \eqn\totalpf{\eqalign{ &\frac{1}{4}\sum_{a,\bar a}\sum_{w
\in \IZ_{2}} \; (-1)^{a+\bar a}\; \bigg\{ \int_0^{\infty} ds
\sqrt{2} \rho(s,w;a,\bar a;\epsilon)
\chi_c\bigg(s,\frac{w+a}{2};\tau,0\bigg)\bigg[ {a \atop 0}\bigg]
\chi_c\bigg(s,\frac{w+\bar a}{2};\bar \tau,0\bigg)\bigg[ {\bar a
\atop 0}\bigg] \cr &+\frac{1}{2}
\chi_d\bigg(\frac{w}{2},\frac{a}{2};\tau,0\bigg)\bigg[ {a \atop
b}\bigg] \chi_d\bigg(\frac{w}{2},\frac{\bar a}{2};\bar
\tau,0\bigg)\bigg[ {\bar a \atop 0}\bigg]\bigg\} \frac{1}{(8\pi^2
\tau_2)^2 \eta^2 \bar \eta^2}\frac{\theta\big[ {a \atop 0 }\big]}
{\eta} \frac{\theta\big[ {\bar a \atop 0 }\big]}{\bar \eta} ~.}}
This sum contains a independent summation over the parameters $a,
\bar a$ accounting for the NS/R-sectors, a summation over the
$U(1)_R$ charges of the ${\cal N}=2$ primaries, and finally either
an integration or a summation over the Casimir eigenvalue of the
coset primaries. An extra minus sign in front of the R-NS or NS-R
sectors accounts for spacetime statistics. This effect is
responsible for the factor $(-1)^{a+\bar a}$.

Tracing over the Hilbert space with an insertion of $(-1)^{J_{\rm GSO}}$ yields
similar results, but with characters having $b=1$. In addition, an extra factor
$(-1)^{\eta a b}$ selects the type IIA or type IIB GSO projection
($\eta=1$ for type IIA and $\eta=0$ for type IIB).
The only subtlety is that since the definition of $b=1$
characters for the coset involves the insertion of $(-1)^{F_c}$,
where $Q_a=2m_t+F_{c}$, a factor
$(-1)^{2  b m_t}=(-1)^{b (w+a)}$ remains explicit.
Finally, an extra factor of $(-1)^{b(a+1)}$
accounts for the superghost contribution to $J_{\rm GSO}$.
Putting everything together and summing over $b,\bar b=0,1$
yields the type II partition sum \appggd .



\listrefs
\end

%% file: tables.tex
%
%
%
\newbox\hdbox%
\newcount\hdrows%
\newcount\multispancount%
\newcount\ncase%
\newcount\ncols
\newcount\nrows%
\newcount\nspan%
\newcount\ntemp%
\newdimen\hdsize%
\newdimen\newhdsize%
\newdimen\parasize%
\newdimen\spreadwidth%
\newdimen\thicksize%
\newdimen\thinsize%
\newdimen\tablewidth%
\newif\ifcentertables%
\newif\ifendsize%
\newif\iffirstrow%
\newif\iftableinfo%
\newtoks\dbt%
\newtoks\hdtks%
\newtoks\savetks%
\newtoks\tableLETtokens%
\newtoks\tabletokens%
\newtoks\widthspec%
%
%
%
%
\tableinfotrue%
\catcode`\@=11
%
%
\def\tstrut{\vrule height4.1ex depth2.2ex width0pt}%
\def\and{\char`\&}
\def\tablerule{\noalign{\hrule height\thinsize depth0pt}}%
\thicksize=1.5pt
\thinsize=0.6pt
\def\thickrule{\noalign{\hrule height\thicksize depth0pt}}%
\def\ctr#1{\hfil\ #1\hfil}%
%
%
%
%
\tablewidth=-\maxdimen%
\spreadwidth=-\maxdimen%
\def\tabskipglue{0pt plus 1fil minus 1fil}%
%
%
\centertablestrue%
%
%
%
%
\parasize=4in%
\gdef\ARGS{########}
\gdef\headerARGS{####}
\def\@mpersand{&}
{\catcode`\|=13
\gdef\letbarzero{\let|0}
\gdef\letbartab{\def|{&&}}%
\gdef\letvbbar{\let\vb|}%
}
{\catcode`\&=4
\def\ampskip{&\omit\hfil&}
\catcode`\&=13
\let&0
\xdef\letampskip{\def&{\ampskip}}%
\gdef\letnovbamp{\let\novb&\let\tab&}
}
\def\begintable{
   \begingroup%
   \catcode`\|=13\letbartab\letvbbar%
   \catcode`\&=13\letampskip\letnovbamp%
   \def\multispan##1{
      \omit \mscount##1%
      \multiply\mscount\tw@\advance\mscount\m@ne%
      \loop\ifnum\mscount>\@ne \sp@n\repeat%
   }
   \def\|{%
      &\omit\widevline&%
   }%
   \ruledtable
}
\long\def\ruledtable#1\endtable{%
%
%
%
   \offinterlineskip
   \tabskip 0pt
   \def\widevline{\vrule width\thicksize}
   \def\endrow{\@mpersand\omit\hfil\crnorm\@mpersand}%
   \def\crthick{\@mpersand\crnorm\thickrule\@mpersand}%
   \def\crthickneg##1{\@mpersand\crnorm\thickrule
          \noalign{{\skip0=##1\vskip-\skip0}}\@mpersand}%
   \def\crnorule{\@mpersand\crnorm\@mpersand}%
   \def\crnoruleneg##1{\@mpersand\crnorm
          \noalign{{\skip0=##1\vskip-\skip0}}\@mpersand}%
   \let\nr=\crnorule
   \def\endtable{\@mpersand\crnorm\thickrule}%
   \let\crnorm=\cr
%
%
   \edef\cr{\@mpersand\crnorm\tablerule\@mpersand}%
   \def\crneg##1{\@mpersand\crnorm\tablerule
          \noalign{{\skip0=##1\vskip-\skip0}}\@mpersand}%
   \let\ctneg=\crthickneg
   \let\nrneg=\crnoruleneg
   \the\tableLETtokens
%
%
   \tabletokens={&#1}
%
%
   \countROWS\tabletokens\into\nrows%
   \countCOLS\tabletokens\into\ncols%
%
%
   \advance\ncols by -1%
   \divide\ncols by 2%
   \advance\nrows by 1%
%
%
   \iftableinfo %
      \immediate\write16{[Nrows=\the\nrows, Ncols=\the\ncols]}%
   \fi%
%
%
   \ifcentertables
      \ifhmode \par\fi
      \line{
      \hss
   \else %
      \hbox{%
   \fi
      \vbox{%
         \makePREAMBLE{\the\ncols}
         \edef\next{\preamble}
         \let\preamble=\next
         \makeTABLE{\preamble}{\tabletokens}
      }
      \ifcentertables \hss}\else }\fi
   \endgroup
   \tablewidth=-\maxdimen
   \spreadwidth=-\maxdimen
}
\def\makeTABLE#1#2{
   {
   \let\ifmath0
   \let\header0
   \let\multispan0
%
%
   \ncase=0%
   \ifdim\tablewidth>-\maxdimen \ncase=1\fi%
   \ifdim\spreadwidth>-\maxdimen \ncase=2\fi%
   \relax
%
   \ifcase\ncase %
      \widthspec={}%
   \or %
      \widthspec=\expandafter{\expandafter t\expandafter o%
                 \the\tablewidth}%
   \else %
      \widthspec=\expandafter{\expandafter s\expandafter p\expandafter r%
                 \expandafter e\expandafter a\expandafter d%
                 \the\spreadwidth}%
   \fi %
   \xdef\next{
      \halign\the\widthspec{%
      #1
      \noalign{\hrule height\thicksize depth0pt}
      \the#2\endtable
%
      }
   }
   }
   \next
}
\def\makePREAMBLE#1{
   \ncols=#1
   \begingroup
   \let\ARGS=0
   \edef\xtp{\widevline\ARGS\tabskip\tabskipglue%
   &\ctr{\ARGS}\tstrut}
   \advance\ncols by -1
   \loop
      \ifnum\ncols>0 %
      \advance\ncols by -1%
      \edef\xtp{\xtp&\vrule width\thinsize\ARGS&\ctr{\ARGS}}%
   \repeat
   \xdef\preamble{\xtp&\widevline\ARGS\tabskip0pt%
   \crnorm}
   \endgroup
}
\def\countROWS#1\into#2{
   \let\countREGISTER=#2%
   \countREGISTER=0%
   \expandafter\ROWcount\the#1\endcount%
}%
\def\ROWcount{%
   \afterassignment\subROWcount\let\next= %
}%
\def\subROWcount{%
   \ifx\next\endcount %
      \let\next=\relax%
   \else%
      \ncase=0%
      \ifx\next\cr %
         \global\advance\countREGISTER by 1%
         \ncase=0%
      \fi%
      \ifx\next\endrow %
         \global\advance\countREGISTER by 1%
         \ncase=0%
      \fi%
      \ifx\next\crthick %
         \global\advance\countREGISTER by 1%
         \ncase=0%
      \fi%
      \ifx\next\crnorule %
         \global\advance\countREGISTER by 1%
         \ncase=0%
      \fi%
      \ifx\next\crthickneg %
         \global\advance\countREGISTER by 1%
         \ncase=0%
      \fi%
      \ifx\next\crnoruleneg %
         \global\advance\countREGISTER by 1%
         \ncase=0%
      \fi%
      \ifx\next\crneg %
         \global\advance\countREGISTER by 1%
         \ncase=0%
      \fi%
      \ifx\next\header %
         \ncase=1%
      \fi%
      \relax%
      \ifcase\ncase %
         \let\next\ROWcount%
      \or %
         \let\next\argROWskip%
      \else %
      \fi%
   \fi%
   \next%
}
\def\counthdROWS#1\into#2{%
\dvr{10}%
   \let\countREGISTER=#2%
   \countREGISTER=0%
\dvr{11}%
\dvr{13}%
   \expandafter\hdROWcount\the#1\endcount%
\dvr{12}%
}%
\def\hdROWcount{%
   \afterassignment\subhdROWcount\let\next= %
}%
\def\subhdROWcount{%
   \ifx\next\endcount %
      \let\next=\relax%
   \else%
      \ncase=0%
      \ifx\next\cr %
         \global\advance\countREGISTER by 1%
         \ncase=0%
      \fi%
      \ifx\next\endrow %
         \global\advance\countREGISTER by 1%
         \ncase=0%
      \fi%
      \ifx\next\crthick %
         \global\advance\countREGISTER by 1%
         \ncase=0%
      \fi%
      \ifx\next\crnorule %
         \global\advance\countREGISTER by 1%
         \ncase=0%
      \fi%
      \ifx\next\header %
         \ncase=1%
      \fi%
\relax%
      \ifcase\ncase %
         \let\next\hdROWcount%
      \or%
         \let\next\arghdROWskip%
      \else %
      \fi%
   \fi%
   \next%
}%
{\catcode`\|=13\letbartab
\gdef\countCOLS#1\into#2{%
   \let\countREGISTER=#2%
   \global\countREGISTER=0%
   \global\multispancount=0%
   \global\firstrowtrue
   \expandafter\COLcount\the#1\endcount%
   \global\advance\countREGISTER by 3%
   \global\advance\countREGISTER by -\multispancount
}%
\gdef\COLcount{%
   \afterassignment\subCOLcount\let\next= %
}%
{\catcode`\&=13%
\gdef\subCOLcount{%
   \ifx\next\endcount %
      \let\next=\relax%
   \else%
      \ncase=0%
      \iffirstrow
         \ifx\next& %
            \global\advance\countREGISTER by 2%
            \ncase=0%
         \fi%
         \ifx\next\span %
            \global\advance\countREGISTER by 1%
            \ncase=0%
         \fi%
         \ifx\next| %
            \global\advance\countREGISTER by 2%
            \ncase=0%
         \fi
         \ifx\next\|
            \global\advance\countREGISTER by 2%
            \ncase=0%
         \fi
         \ifx\next\multispan
            \ncase=1%
            \global\advance\multispancount by 1%
         \fi
         \ifx\next\header
            \ncase=2%
         \fi
         \ifx\next\cr       \global\firstrowfalse \fi
         \ifx\next\endrow   \global\firstrowfalse \fi
         \ifx\next\crthick  \global\firstrowfalse \fi
         \ifx\next\crnorule \global\firstrowfalse \fi
         \ifx\next\crnoruleneg \global\firstrowfalse \fi
         \ifx\next\crthickneg  \global\firstrowfalse \fi
         \ifx\next\crneg       \global\firstrowfalse \fi
      \fi
\relax
      \ifcase\ncase %
         \let\next\COLcount%
      \or %
         \let\next\spancount%
      \or %
         \let\next\argCOLskip%
      \else %
      \fi %
   \fi%
   \next%
}%
\gdef\argROWskip#1{%
   \let\next\ROWcount \next%
}
\gdef\arghdROWskip#1{%
   \let\next\ROWcount \next%
}
\gdef\argCOLskip#1{%
   \let\next\COLcount \next%
}
}
}
\def\spancount#1{
   \nspan=#1\multiply\nspan by 2\advance\nspan by -1%
   \global\advance \countREGISTER by \nspan
   \let\next\COLcount \next}%
\def\dvr#1{\relax}%
\def\header#1{%
\dvr{1}{\let\cr=\@mpersand%
\hdtks={#1}%
\counthdROWS\hdtks\into\hdrows%
\advance\hdrows by 1%
\ifnum\hdrows=0 \hdrows=1 \fi%
\dvr{5}\makehdPREAMBLE{\the\hdrows}%
\dvr{6}\getHDdimen{#1}%
{\parindent=0pt\hsize=\hdsize{\let\ifmath0%
\xdef\next{\valign{\headerpreamble #1\crnorm}}}\dvr{7}\next\dvr{8}%
}%
}\dvr{2}}
\def\makehdPREAMBLE#1{
\dvr{3}%
\hdrows=#1
{
\let\headerARGS=0%
\let\cr=\crnorm%
\edef\xtp{\vfil\hfil\hbox{\headerARGS}\hfil\vfil}%
\advance\hdrows by -1
\loop
\ifnum\hdrows>0%
\advance\hdrows by -1%
\edef\xtp{\xtp&\vfil\hfil\hbox{\headerARGS}\hfil\vfil}%
\repeat%
\xdef\headerpreamble{\xtp\crcr}%
}
\dvr{4}}
\def\getHDdimen#1{%
\hdsize=0pt%
\getsize#1\cr\end\cr%
}
\def\getsize#1\cr{%
\endsizefalse\savetks={#1}%
\expandafter\lookend\the\savetks\cr%
\relax \ifendsize \let\next\relax \else%
\setbox\hdbox=\hbox{#1}\newhdsize=1.0\wd\hdbox%
\ifdim\newhdsize>\hdsize \hdsize=\newhdsize \fi%
\let\next\getsize \fi%
\next%
}%
\def\lookend{\afterassignment\sublookend\let\looknext= }%
\def\sublookend{\relax%
\ifx\looknext\cr %
\let\looknext\relax \else %
   \relax
   \ifx\looknext\end \global\endsizetrue \fi%
   \let\looknext=\lookend%
    \fi \looknext%
}%
%
%
\def\tablelet#1{%
   \tableLETtokens=\expandafter{\the\tableLETtokens #1}%
}%
\catcode`\@=12
%